\documentclass[journal]{IEEEtran}
\usepackage{amsmath,amsfonts,amsthm}
\usepackage[ruled,commentsnumbered,linesnumbered]{algorithm2e}
\usepackage{array}
\usepackage{textcomp}
\usepackage[dvipsnames]{xcolor}
\usepackage{stfloats}
\ifCLASSOPTIONcompsoc
\usepackage[caption=false,font=normalsize,labelfont=sf,textfont=sf]{subfig}
\else
\usepackage[caption=false,font=footnotesize]{subfig}
\fi
\usepackage{url}
\usepackage{verbatim}
\usepackage{graphicx}
\usepackage{cite}
\usepackage[colorlinks,linkcolor=BrickRed,citecolor=Aquamarine,urlcolor=cyan]{hyperref} 
\usepackage{makecell}
\usepackage{multirow}
\usepackage{color}
\usepackage{booktabs}
\usepackage{amsmath}
\usepackage{colortbl}
\usepackage{bbm}
\usepackage{textcomp}
\usepackage{xcolor}

\allowdisplaybreaks[4]
\usepackage{amssymb}
\usepackage{threeparttable}

\theoremstyle{remark}
\newtheorem{remark}{Remark}
\newtheorem{proposition}{Proposition}
\newtheorem{lemma}{Lemma}

\newtheorem{theorem}{Theorem}

\definecolor{lightblue}{rgb}{0.93,0.95,1.0}

\begin{document}

\title{Mixed-Precision Quantization: Make the Best Use of Bits Where They Matter Most}
\author{Yiming Fang,~\IEEEmembership{Graduate Student Member,~IEEE}, 
    Li Chen,~\IEEEmembership{Senior Member,~IEEE},
    \\
   Yunfei Chen,~\IEEEmembership{Senior Member,~IEEE},
    Weidong Wang,
    and Changsheng You,~\IEEEmembership{Member,~IEEE}
    
    \thanks{Yiming Fang, Li Chen, and Weidong Wang are with the CAS Key Laboratory of Wireless-Optical Communications, University of Science and Technology of China, Hefei 230027, China (e-mail: fym1219@mail.ustc.edu.cn; chenli87@ustc.edu.cn; wdwang@ustc.edu.cn).}
    \thanks{Yunfei Chen is with the Department of Engineering, University of Durham, Durham, UK, DH1 3LE (e-mail: Yunfei.Chen@durham.ac.uk).}
    \thanks{Changsheng You is with the Department of Electronic and Electrical Engineering, Southern University of Science and Technology (SUSTech), Shenzhen 518055, China (e-mail: youcs@sustech.edu.cn).}
}



\maketitle
\begin{abstract}
Mixed-precision quantization offers superior performance to fixed-precision quantization. It has been widely used in signal processing, communication systems, and machine learning. In mixed-precision quantization, bit allocation is essential. Hence, in this paper, we propose a new bit allocation framework for mixed-precision quantization from a search perspective. First, we formulate a general bit allocation problem for mixed-precision quantization. Then we introduce the penalized particle swarm optimization (PPSO) algorithm to address the integer consumption constraint. To improve efficiency and avoid iterations on infeasible solutions within the PPSO algorithm, a greedy criterion particle swarm optimization (GC-PSO) algorithm is proposed. The corresponding convergence analysis is derived based on dynamical system theory. Furthermore, we apply the above framework to some specific classic fields, i.e., finite impulse response (FIR) filters, receivers, and gradient descent. Numerical examples in each application underscore the superiority of the proposed framework to the existing algorithms.
\end{abstract}


\section{Introduction}
\label{sec: intro}
\IEEEPARstart{L}{arge-scale} signal processing, communications, and machine learning (ML) have garnered increasing attention in recent years \cite{10379539,7091863,9165233}. In many of these applications, the complexity and overhead are unbearable due to the substantial number of antennas or data volumes \cite{dean2012large}. One approach to alleviating these complexities and bottlenecks is quantization.

Conventional quantization uses fixed uniform low-precision quantization, and has been well-studied in signal processing, communications, and ML \cite{6094235,7938758,7478040,7307134,7439790,9043731}. For example, in signal processing, low-precision quantization has been applied to finite impulse response (FIR) filter design \cite{6094235}, subspace estimation \cite{7938758}, and direction-of-arrival (DOA) estimation \cite{7478040}. Moreover, the authors in \cite{7307134} and \cite{7439790} employed low-precision quantization or analog-to-digital converters (ADCs) for massive multiple-input-multiple-output (MIMO) communication and channel estimation, respectively. In addition to signal processing and communication systems, low-precision quantization has enabled neural network (NN) compression and acceleration \cite{9043731}. 

The above scheme applies the same quantization bits to all the inputs. Such a uniform bit allocation can be sub-optimal, since different inputs exhibit different redundancies, such as magnitude, and sensitivity to bits, and contribute differently to the final performance. Therefore, if we can utilize mixed-precision quantization, i.e., different inputs allocated with different quantization bits, it is possible to achieve a more effective balance between performance and complexity. This can be enabled by some hardware accelerators \cite{9043731, sharma2018bit} that support mixed-precision computation.

Mixed-precision quantization has applications in signal processing, communications, and ML. Specifically, in signal processing, the Cramér-Rao bound of DOA estimation based on mixed-ADC was analyzed in \cite{10541876}. Furthermore, the authors in \cite{10185129} applied mixed-precision quantization to enhance signal detection with a bandwidth-constrained distributed radar system. For communication systems, the authors in \cite{7562390} proposed an advanced detector for mixed-ADC massive MIMO systems. Moreover, the performance analysis of mixed-ADC was presented in \cite{7437384}. In the context of ML, the authors in \cite{dettmers2022gpt3} reduced the large language model overhead by assigning more bits for emergent features with large magnitudes and fewer bits for those with small magnitudes.

The above studies employ heuristic mixed-precision quantization. For instance, inputs with large magnitudes are assigned more bits, while entries with small magnitudes are assigned fewer bits. A more efficient approach to mixed-precision quantization is to determine bit allocation through optimization. To this end, some works have formulated different optimization problems with integer consumption constraints to determine the bit allocation. For example, the authors in \cite{8017448} and \cite{9966648} provided the bit allocation schemes by minimizing mean square error in millimeter wave and cell-free MIMO systems with precision-adaptive ADC. Moreover, mixed-precision Bayesian parameter estimation was studied in \cite{9571074}. Besides, the authors in \cite{10508306} proposed a low-complexity harmony search (HS)-based algorithm for bit allocation in cell-free massive MIMO systems. In NN compression, the authors in \cite{Chen_2021_ICCV} formulated mixed-precision quantization as a discrete constrained optimization problem to determine the bit allocation for tensors across layers. Bit allocation for activation was further addressed in \cite{9399174} as an optimization problem. For wireless federated learning, the authors in \cite{10091800} minimized the convergence rate upper bound under a quantization resource budget. However, these works transform the original optimization problems into a convex optimization by relaxing constraints or approximating original objective functions with Taylor series expansion, which can result in sub-optimal performance. The optimal approach to mixed-precision quantization is to determine bit allocation by searching the entire feasible space under integer consumption constraints. The corresponding challenge lies in developing efficient search algorithms, as the complexity of brute-force search is prohibitively high. To the best of our knowledge, determining bit allocation for mixed-precision quantization through efficient searching under integer consumption constraints remains an open problem.


To achieve this, in this paper, we propose a bit allocation framework for mixed-precision quantization from a search perspective. Specifically, we first formulate a general bit allocation problem for mixed-precision quantization. Particle swarm optimization (PSO) is a promising low-complexity algorithm for achieving near-optimal performance\footnote{Other meta-heuristic algorithms, such as ant-colony optimization (ACO), genetic algorithm (GA), and simulated annealing (SA), exhibit specific limitations in the context of bit allocation. Specifically, ACO is primarily developed for discrete pathfinding problems, such as the shortest path problem, and is less effective for functional optimization tasks like fronthaul bit allocation \cite{10508306}. GA generates new solutions by combining pairs of parent solutions, which can lead to redundancy and overlapping solutions, reducing the diversity of the population and potentially degrading performance \cite{nojima2005effects}. SA operates on a single solution and lacks population-based search, resulting in limited exploration capability and suboptimal convergence behavior compared to algorithms with directional or greedy updates \cite{panda2018performance}.}. However, the conventional PSO algorithm cannot be applied directly to integer-constrained searching problems. Therefore, two PSO-based algorithms are proposed to address the general mixed-precision quantization searching problem. Furthermore, we extend the above design to different classic fields, i.e., FIR filters, receivers, and gradient descent (GD). Finally, numerical examples demonstrate the superiority of the proposed search framework. Our main contributions are summarized as follows. 
\begin{itemize}
    \item \textbf{Bit Allocation Framework for Mixed-Precision Quantization.} We propose a bit allocation framework for mixed-precision quantization from a search perspective. Specifically, we formulate a general bit allocation problem for mixed-precision quantization. To address the integer consumption constraint, we introduce the penalized PSO (PPSO) algorithm. Then, we propose a greedy criterion PSO (GC-PSO) algorithm to reduce iterations on these infeasible solutions in the PPSO algorithm. Moreover, the corresponding convergence analysis is derived based on dynamical system theory.
    \item \textbf{Mixed-Precision FIR Filter Design.} The bit allocation framework is applied to the FIR filter design based on a mixed-precision minimax approximation problem. Moreover, we present low-complexity solutions to the minimum mean square error (MMSE) problem under fixed-point quantization and floating-point quantization to find the bit allocation of the FIR filter, respectively. Numerical examples demonstrate that our algorithms outperform the best methods in \cite{kodek2005telescoping, 8307077}.
    \item \textbf{Receiver with Precision-Adaptive ADC.} We apply the bit allocation framework to receivers in massive MIMO systems with precision-adaptive ADC architecture. Specifically, a sum achievable rate maximization problem with precision-adaptive ADC is addressed to determine the bit allocation. Simulation results indicate the superiority of our proposed algorithms compared to the method presented in \cite{8017448,10508306}.
    \item \textbf{Mixed-Precision Gradient Descent.} The bit allocation framework is utilized in a distributed GD scenario involving a server and a worker. In particular, we solve a minimum loss function problem under a total quantization bits constraint at each iteration to ascertain bit allocation. Numerical results reveal that our proposed algorithms demonstrate improved convergence compared to fixed-precision quantization methods using the least squares problem and logistic regression for binary classification as examples.
\end{itemize}

\textit{Organization:} Section \ref{sec: mixed_precision_quantization} provides a search framework for mixed-precision quantization. In Section \ref{sec: fir}, we apply the proposed algorithms to the FIR filter design. Section \ref{sec: adc} applies the proposed algorithms to receivers in massive MIMO systems with precision-adaptive ADC architecture. We use the proposed algorithms to address the quantization bit allocation for quantized GD in Section \ref{sec: GD}. The conclusions are provided in Section \ref{sec: con}.

\textit{Notation:} Bold uppercase letters denote matrices and bold lowercase letters denote vectors. For a matrix $\bf A$, ${\bf A}^T$, ${\bf A}^H$ and ${\bf A}^{-1}$ denote the transpose, the Hermitian transpose and inverse of ${\bf A}$, respectively. ${a}_{ij}$ denotes $(i,j)$-th entry of ${\bf A}$. $\mathrm{tr}(\bf A)$ denotes the trace of matrix ${\bf A}$. $\mathrm{diag}(\bf A)$ denotes the matrix of the diagonal elements of matrix ${\bf A}$. $\mathbb{E}\{\bf A\}$ denotes the expectation of $\bf A$. For a vector $\bf a$, $\left\| \bf a \right\|_2$ denotes its Euclidean norm. The notations $\mathbb{N}$, $\mathbb{Z}$, $\mathbb{Z}_+$, $\mathbb{R}$, and $\mathbb{C}$ represent the sets of nature numbers, integer numbers, positive integer numbers, real numbers, and complex numbers, respectively. $\# \mathbb{B}$ is the number of elements in set $\mathbb{B}$. $\lceil x \rceil$ and $\lfloor x \rfloor$ represent the smallest integer more than $x$ and the largest integer no more than $x$, respectively.

\section{Bit Allocation Framework for Mixed-Precision Quantization}
\label{sec: mixed_precision_quantization}
\subsection{Problem Formulation}
 \begin{figure}[t]
     \centering
     \includegraphics[width=0.35\textwidth]{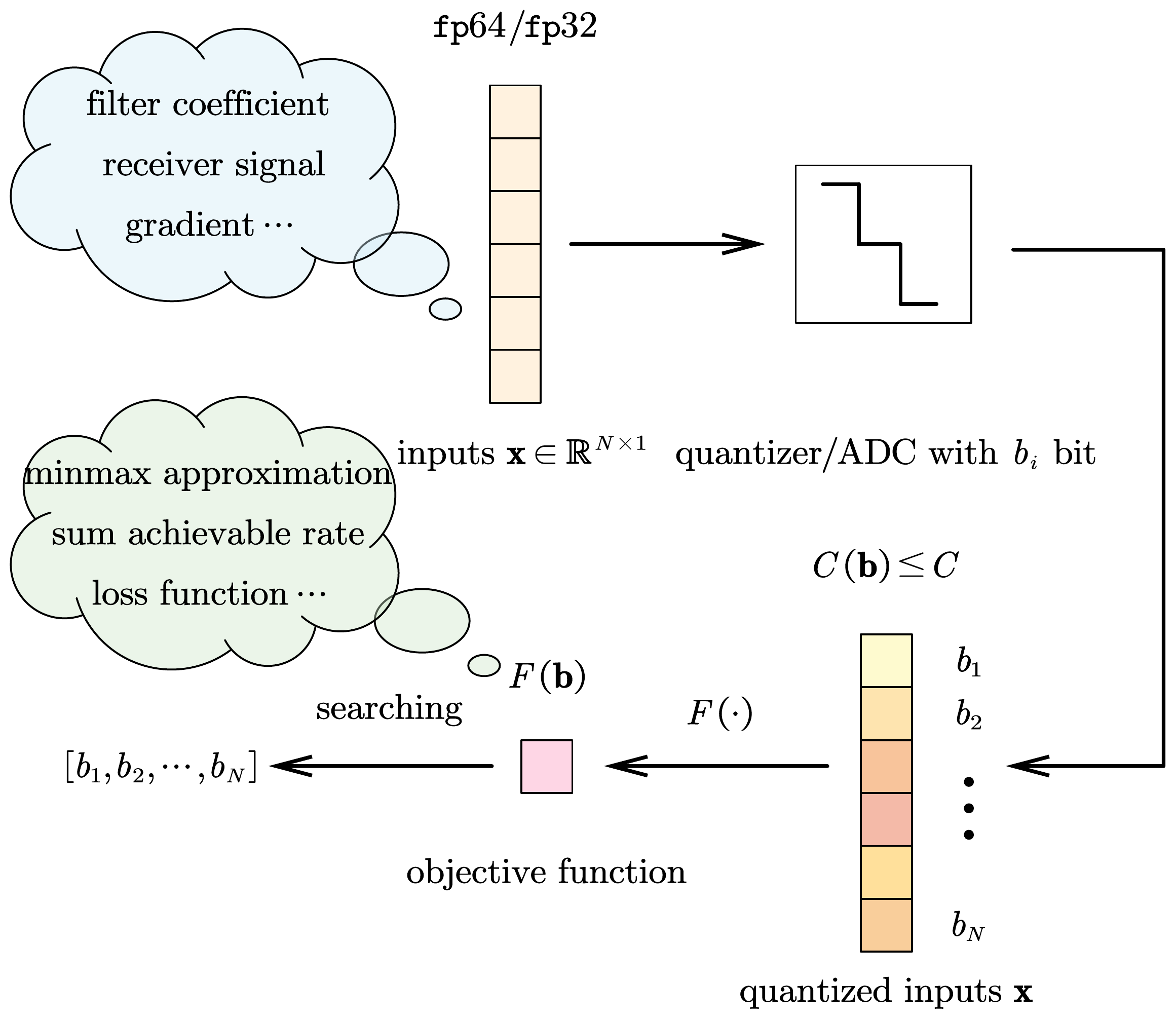}
     \caption{The illustration of bit allocation framework for mixed-precision quantization. Inputs such as filter coefficients, base station receiver signals, and gradients of the GD algorithm are always quantized due to hardware limitations or communication costs. To make the best use of bits, mixed-precision quantization has become a widely adopted approach. In the figure, different colors correspond to different numbers of quantization bits. Using the quantized inputs with different bits, we can obtain different objective functions such as minimax approximate, sum achievable rate, and loss under the total consumption constraint. Finally, the bit allocation can be determined by searching.}
     \label{fig: general_problem}
 \end{figure}
 
In this section, we propose a bit allocation framework for mixed-precision quantization from a search perspective as illustrated in Fig. \ref{fig: general_problem}. Specifically, given the quantization bit sequence ${\bf b} = \{b_n\}_{n=1}^{N}$, we can formulate the bit allocation problem for mixed-precision quantization as follows:
\begin{subequations}
\begin{align} 
{\left ({{{\mathcal{P}_1}} }\right)}~{\underset{\{ b_n \}_{n=1}^{N}}{\min}  }&~{F\left({\bf b}\right)}  \label{eq: obj_ge}\\ 
{{\text {s.t.}}}~~\,&~{C\left({\bf b}\right)\leq C(\bar{b}),}  \label{eq: consumption}\\
&~\ b_n \in \mathbb{B},~n = 1, 2, \dots, N,\label{eq: b_set}
\end{align}
\end{subequations}
where $F\left({\bf b}\right)$ is a general objective metric function of $\bf b$, such as MMSE, minimax, and cross-entropy loss, $C\left({\bf b}\right)$ represents the consumption function, $\bar{b}$ in \eqref{eq: consumption} is the average number of quantization bits and nonempty constraint $\mathbb{B}\subseteq \mathbb{Z} $ in \eqref{eq: b_set} is the set of allowable quantization bit values. Constraint \eqref{eq: consumption} limits the total consumption of quantization bits, while constraint \eqref{eq: b_set} defines the allowable values for the quantization bits.

Note that solving problem $\left(\mathcal{P}_1\right)$ is challenging. Since $\{b_n\}_{n=1}^{N}$ are non-negative integers, problem $\left(\mathcal{P}_1\right)$ is actually a searching problem with integer programming constraint, which is NP-hard \cite{schrijver1998theory}. Further, if brute-force search is utilized to solve problem $\left(\mathcal{P}_1\right)$, the time complexity will be $\mathcal{O} \left( \left( \#\mathbb{B} \right) ^N \right)$, which is infeasible for even short quantization bit sequence. Although the classic PSO algorithm can be applied to address \eqref{eq: obj_ge}, it cannot be directly used in the integer-constrained searching problem. Consequently, we propose two PSO-based algorithms to efficiently solve problem $\left(\mathcal{P}_1\right)$ in the following.

In addition to problem $\left({{\mathcal{P}_1}}\right)$, minimization of the total consumption problems can also be considered. The general problem of minimizing the total consumption with mixed-precision quantization can be formulated as follows:
    \begin{align*} 
    {\left ({{{\mathcal{P}_1^{'}}} }\right)}~{\underset{\{ b_n \}_{n=1}^{N}}{\min}  }&~{C\left({\bf b}\right)} \\ 
    {{\text {s.t.}}}~~\,&~{F\left({\bf b}\right)\leq F(\bar{b}),}~\eqref{eq: b_set}.
    \end{align*}
Problems $\left({\mathcal{P}_1}\right)$ and $\left({\mathcal{P}_1^{'}}\right)$ exhibit a dual structure in terms of their objective and constraint formulations. In particular, the roles of the objective function $F\left({\bf b}\right)$ and the consumption function $C\left({\bf b}\right)$ are interchanged across the two formulations. In other words, the algorithms for solving problem $\left({\mathcal{P}_1}\right)$ can address problem $\left({\mathcal{P}_1^{'}}\right)$. Therefore, for simplicity, the remainder of this paper will focus on problem $\left({{\mathcal{P}_1}}\right)$.


\subsection{Penalized PSO Algorithm}
\label{sec: ppso}
First, we determine the fitness/objective function for the PPSO algorithm. Specifically, to handle the inequality in \eqref{eq: consumption}, the problem $\left(\mathcal{P}_1\right)$ can be transformed into an unconstrained form as
\begin{align}
    {\left ({{{\mathcal{P}_{1.1}}} }\right)}~{\underset{\{ b_n \}_{n=1}^{N}}{\min}  }~{F\left({\bf b}\right) + \lambda \max \left( 0, C\left({\bf b}\right)- C(\bar{b})\right)},  \label{eq: obj_p11}
\end{align}
where $\lambda>0$ is a penalty parameter, $b_n\in \mathbb{B}$. Thus, \eqref{eq: obj_p11} serves as the fitness function for the PPSO algorithm.

Then, in the PPSO algorithm, a swarm of particles explore the solution space, where each particle's position represents a potential solution to the optimization problem. We map the optimization target, i.e., quantization bit sequence, to the position of each particle. The specific PPSO model of $N_{\rm pop}$ particles is defined as follows. For $(\rm it)$-th iteration and $i$-th particle, the model is given by:
\begin{align}
    \mathbf{v}_{i}^{\mathrm{it}+1}&=w\mathbf{v}_{i}^{\mathrm{it}}+c_1{r}_1\left( \mathbf{b}_{p,i}^{\mathrm{best}}-\mathbf{b}_{i}^{\mathrm{it}} \right) +c_2{r}_2\left( \mathbf{b}_{\mathrm{g}}^{\mathrm{best}}-\mathbf{b}_{i}^{\mathrm{it}} \right),\label{eq: velocity}\\
\mathbf{b}_{i}^{\mathrm{it}+1}&=\mathbf{b}_{i}^{\mathrm{it}}+\mathtt{round}\left( \mathbf{v}_{i}^{\mathrm{it}+1} \right),\label{eq: poistion}
\end{align}
where $\mathbf{b}_{p,i}^{\mathrm{best}}$ and $\mathbf{b}_{\mathrm{g}}^{\mathrm{best}}$ are the personal best position of the $i$-th particle and the global best position until $(\rm it)$-th iteration, respectively. $\mathbf{b}_{i}^{\mathrm{it}}$ and $\mathbf{v}_{i}^{\mathrm{it}}$ are the position and velocity of $i$-th particle, respectively. Here, $\omega$ is the inertia weight, $c_1$ and $c_2$ are the cognitive and social acceleration coefficients, respectively, and $r_1$ and $r_2$ are uniform random variables satisfying $\mathcal{U}(0,1)$. Moreover, $\mathtt{round}$ is the rounding function that ensures each particle's position is integer.

\begin{algorithm}[t]
\label{alg: PPSO}
    \SetAlgoNlRelativeSize{0}
    \SetAlgoNlRelativeSize{-1}
    \caption{PPSO Algorithm for General Mixed-Precision Quantization Problem}
    \KwIn{$\Bar{b}$, $N_{\rm pop}$, $w_{\rm max}$, $w_{\rm min}$, $\mathbb{B}$, $v_{\rm max}$, $v_{\rm min}$, $c_{1, \mathrm{max} }$, $c_{1, \mathrm{min} }$, $c_{2, \mathrm{max} }$, $c_{2, \mathrm{min} }$, $I_{\rm iter}$}
    \KwOut{The optimal bit allocation ${\bf b}_{\rm opt}$}

    Set ${\bf b}_{\rm g}^{\rm best}\in \mathbb{Z}$ randomly in range $\mathbb{B}$
    
    Set $C_{\rm g}^{\rm best}$ to $\infty$
    
    \For{$i = 1:N_{\rm pop}$}{
        Initialize ${\bf b}_i$ using \eqref{eq: b_init}
        
        Initialize ${\bf v}_i$ randomly in range $[v_{\rm min}, v_{\rm max}]$
        
        Compute cost using \eqref{eq: obj_p11}
        
        Update ${\bf b}_{p,i}^{\rm best}$, $C_{i}^{\rm best}$, ${\bf b}_{\rm g}^{\rm best}$ and $C_{\rm g}^{\rm best}$
    }
    \For{${\rm it} = 1:I_{\rm iter}$}{
        Using \eqref{eq: omega}, \eqref{eq: c1}, and \eqref{eq: c2}

        \For{$i = 1:N_{\rm pop}$}{
            Compute ${\bf v}_{i}$ using \eqref{eq: velocity}, Apply velocity limits
            
            Update ${\bf b}_i$ using \eqref{eq: poistion}, Apply position limits
            
            Compute cost using \eqref{eq: obj_p11}
            
            Update ${\bf b}_{p,i}^{\rm best}$, $C_{i}^{\rm best}$, ${\bf b}_{\rm g}^{\rm best}$ and $C_{\rm g}^{\rm best}$
        }
        
                

            
    }
    return ${\bf b}_{\rm opt}={\bf b}_{\rm g}^{\rm best}$
\end{algorithm}

Furthermore, the initial position of the $i$-th particle is determined based on the average quantization bit $\Bar{b}$, which is calculated as follows:
\begin{equation}
    \label{eq: b_init}
    \mathbf{b}_{i} = \left[ \bar{b},\bar{b},\cdots ,\bar{b} \right]^T \in \mathbb{Z} ^{N\times 1 }.
\end{equation}
This initialization approach can achieve better convergence than random initialization.

To further mitigate the risk of being trapped in local minima, we adopt a time-varying hyper-parameter updating technique for $\omega$, $c_1$, and $c_2$ \cite{785511,1304846}. Specifically, for the $(\rm it)$-th iteration, we define:
\begin{align}
        w &= w_{\rm max} - \left(w_{\rm max}-w_{\rm min}\right)\left({\rm it}/I_{\rm iter}\right), \label{eq: omega}\\
        c_1 &= c_{1, \mathrm{max} } + \left(c_{1, \mathrm{min} }-c_{1, \mathrm{max} }\right)\left({\rm it}/I_{\rm iter}\right), \label{eq: c1}\\        
        c_2 &= c_{2, \mathrm{min} } + \left(c_{2, \mathrm{max} }-c_{2, \mathrm{min} }\right)\left({\rm it}/I_{\rm iter}\right), \label{eq: c2}
\end{align}
where $w_{\rm max}$ and $w_{\rm min}$ are the initial and final values of the inertia weight, respectively. $c_{1, \mathrm{max} }$, $c_{1, \mathrm{min}}$, $c_{2, \mathrm{max} }$ and $c_{2, \mathrm{min}}$ are the initial and final acceleration coefficients, respectively\footnote{The specific hyperparameters in \eqref{eq: omega}, \eqref{eq: c1}, and \eqref{eq: c2} have been studied in detail in \cite{785511,1304846,9680690} and can be adopted following the configurations provided in these references.}. $I_{\rm iter}$ represents the maximum number of iterations. 

The overall procedure of the PPSO algorithm is summarized in \textit{Algorithm} \ref{alg: PPSO}. The time complexity of \textit{Algorithm} \ref{alg: PPSO} will be detailed in the next section, based on the specific application. Notably, during the initial iteration stage, many particles may fall into the infeasible solution space, causing the PPSO algorithm to spend a lot of iterations repairing these infeasible solutions rather than exploring better feasible alternatives. For a better search of the feasible solution space, next we propose a GC-PSO algorithm that operates without a penalty term in the following subsection.

\subsection{Greedy Criterion PSO Algorithm}
\label{sec: gc-pso}
Compared with the PPSO algorithm, the GC-PSO algorithm mainly has two differences. First, the fitness function of the GC-PSO algorithm is the original objective function \eqref{eq: obj_ge} without the need for a penalty term. Second, after updating the particle positions using \eqref{eq: poistion}, the following greedy criterion procedure is applied for each particle:
\subsubsection{Sensitivity} We define a metric known as the sensitivity of quantization noise to evaluate which quantization bit needs to be changed during a single cycle. Specifically, the sensitivity of quantization noise when the $j$-th quantization bit is modified can be expressed as
\begin{align}\label{eq: senti}
    S_j = F\left(\hat{\bf b}\right) - F\left({\bf b}\right),
\end{align}
where ${\bf b}=[b_1,\cdots,b_{j},\cdots,b_{N}]^T$ is the original quantization bit sequence, and $\hat{\bf b}=[b_1,\cdots,\Hat{b}_{j},\cdots,b_{N}]^T$ is the quantization bit sequence after modifying the $j$-th quantization bit.
\subsubsection{Update Criterion} If the constraint \eqref{eq: consumption} is violated, i.e., the total bit number exceeds the maximum allowable limit, we identify the minimum sensitivity in \eqref{eq: senti} for $j=1,2,\cdots, N$ and reduce the corresponding quantization bit in each cycle until the constraint \eqref{eq: consumption} is satisfied.

\begin{algorithm}[t]
\label{alg: GC-PSO}
    \SetAlgoNlRelativeSize{0}
    \SetAlgoNlRelativeSize{-1}
    \caption{Function for Bit Adjustment and Constraint Satisfaction with Greedy Criterion}
    \KwIn{$\Bar{b}$, $\mathbb{B}$, ${\bf b}$}
    \KwOut{The bit allocation ${\bf b}_{\rm g}$}

    Repair the quantization bit sequence $\bf b$ into range $\mathbb{B}$ and ensure they are integers by rounding

    Compute the total consumption $C\left({\bf b}\right)$, and the maximum consumption $C(\bar{b})$

    \If{$C\left({\bf b}\right)>C(\bar{b})$}{
    
    Scale the quantization bit sequence by ${\bf b} = \mathtt{round}({\bf b}\times \frac{C(\bar{b})}{C\left({\bf b}\right)})$

    Recompute the total consumption $C\left({\bf b}\right)$
    
    Initialize sensitivity vector with zero vector
     
    \While{$C\left({\bf b}\right)>C(\bar{b})$}{
        Compute current $F\left({\bf b}\right)$ with $\bf b$
        
        \For{$j = 1:N$}{
         \eIf{$b_j > \min{\mathbb{\left(B\right)}}$}{
                Create temporary bit vector $\hat{\bf b} = {\bf b}$
                
                Set $\hat{b}_j = \hat{b}_j - 1$\;
                
                Compute new $F\left(\hat{\bf b}\right)$ with $\hat{\bf b}$
                
                Calculate the sensitivity using \eqref{eq: senti}
                }
        {
            Set the sensitivity as $\infty$
        }
        }
        Sort the sensitivity vector ${\bf S}=[S_1,\cdots,S_N]$, and find the lowest one as $j^*$

        Update the bit assignment by $b_{j^*} = \hat{b}_{j^*}$

        Recompute the total consumption $C\left({\bf b}\right)$

    }
    }
    return ${\bf b}_{\rm g} = \bf b$
\end{algorithm}

 \begin{figure}[t]
     \centering
     \includegraphics[width=0.35\textwidth]{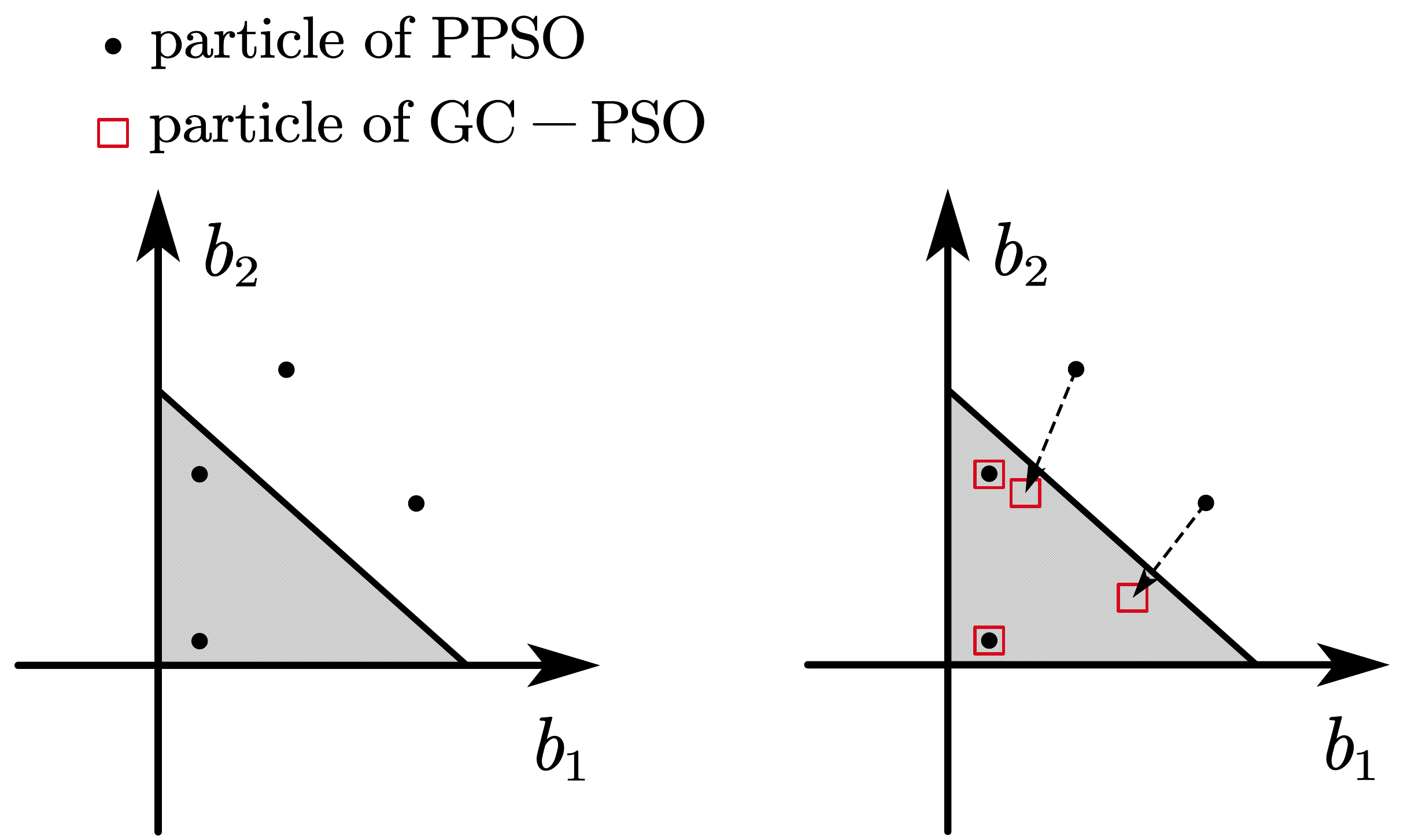}
     \caption{Toy of particle distribution for PPSO and GC-PSO algorithms at an iteration with $N=2$ and number of particles $N_{\rm pop}=4$. The shaded area denotes the feasible solution space. Some particles of the PPSO algorithm (represented by black dots in the figure) may lie outside the feasible solution space. The GC-PSO algorithm can repair these particles, and particles of the GC-PSO algorithm (represented by red squares in the figure) consistently remain within this space.}
     \label{fig: ppso_vs_gcpso}
 \end{figure}
 
The remaining processes of the GC-PSO algorithm are similar to those of the PPSO algorithm and are therefore omitted for brevity. The overall procedure of the greedy criterion for bit adjustment and constraint satisfaction is summarized in \textit{Algorithm} \ref{alg: GC-PSO}. Furthermore, the specific time complexity of the GC-PSO algorithm will be provided in the following section. Compared to the PPSO algorithm, the time complexity of the GC-PSO algorithm is higher due to the presence of a while loop in \textit{Algorithm} \ref{alg: GC-PSO}. Nevertheless, the GC-PSO algorithm demonstrates superior performance based on numerical examples. Fig. \ref{fig: ppso_vs_gcpso} compares a toy of particle distribution for PPSO and GC-PSO algorithms at an iteration with $N=2$ and number of particles $N_{\rm pop}=4$.

\subsection{Convergence Analysis}
Similar to the most theoretical convergence analysis of PSO and its variants \cite{6819057,985692,TRELEA2003317}, the convergence properties of the PPSO and GC-PSO algorithms are analyzed under a deterministic implementation. Specifically, we reduce \eqref{eq: velocity} and \eqref{eq: poistion}, i.e., the velocity and position evolutionary equations of the PPSO and GC-PSO algorithms, to the one-dimension deterministic case and omit subscript $i$, yielding:
\begin{align}
    v^{\mathrm{it}+1}&=wv^{\mathrm{it}}+\frac{c_1}{2}\left( b_{p}^{\mathrm{best}}-b^{\mathrm{it}} \right) +\frac{c_2}{2}\left( b_{\mathrm{g}}^{\mathrm{best}}-b^{\mathrm{it}} \right) , \label{eq: ve_one} \\
    b^{\mathrm{it}+1}&=b^{\mathrm{it}}+\mathtt{round}\left( v^{\mathrm{it}+1} \right), \label{eq: po_one}
\end{align}
where $w$, $c_1$, $c_2$ are set to be constant, and $\frac{1}{2}$ is the expected value of $r_1$ and $r_2$. 

Compared to existing convergence analyses of PSO and its variants, analyzing the PPSO and GC-PSO algorithms is challenging due to the non-linear and non-differentiable $\mathtt{round}$ function. In the following theorem, we derive the convergence conditions for PPSO and GC-PSO algorithms from the perspective of dynamical systems.

\begin{theorem}
    \label{the: con}
    Given $0<w<c+1$ and the largest eigenvalue of matrix $\mathbf{P}$ $\lambda_{\max} ({\bf P})<\frac{1}{2\sqrt{c^2+w^2}}$, the dynamic system described by \eqref{eq: ve_one} and \eqref{eq: po_one} converges to an equilibrium point, where
    \begin{align}
        \mathbf{P}=\left[ \begin{matrix}
	\frac{c+c^2+w^2}{2c\left( 1+c-w \right)}&		\frac{-c^2+w-w^2}{2c\left( 1+c-w \right)}\\
	\frac{-c^2+w-w^2}{2c\left( 1+c-w \right)}&		\frac{1+c+c^2-2w+w^2}{2c\left( 1+c-w \right)}\\
\end{matrix} \right],~ c=\frac{c_1+c_2}{2}. \notag
    \end{align}
\end{theorem}
\begin{IEEEproof}
    The proof is available in Appendix \ref{app: the}.
\end{IEEEproof}
\textit{Theorem} \ref{the: con} presents a sufficient condition for the convergence of the proposed PPSO and GC-PSO algorithms. It is derived by modeling the simplified deterministic dynamics of particle motion as a discrete-time system. The result shows that under $0<w<c+1$ and $\lambda_{\max} ({\bf P})<\frac{1}{2\sqrt{c^2+w^2}}$, the particles will converge to a stable equilibrium point, despite the non-linear rounding operation. It should also be pointed out that the convergence condition is derived based on the worst-case analysis, which may be conservative.


\section{Mixed-Precision Application \uppercase\expandafter{\romannumeral1}: FIR Filter}
\label{sec: fir}
In this section, we consider the application of the proposed algorithms to FIR filter design. First, a mixed-precision minimax approximation problem is formulated. Then, we apply the proposed algorithms and present low-complexity solutions. Finally, we present numerical results to demonstrate the superiority of the proposed algorithms.

\subsection{Problem Statement}

 
We consider an $N$-tap linear-phase direct FIR filter with real-valued impulse response ${\bf h} = \{h[n]\}_{n=0}^{N-1}$. The corresponding frequency response is given by 
\begin{align}
        H\left( e^{{j}\omega} \right) &= \sum_{n=0}^{N-1}{h\left[ n \right] e^{-{{j}}\omega n}} \\
    &= H\left( \omega \right) e^{{j}\left( \frac{L\pi}{2}-\frac{N-1}{2}\omega \right)},
\end{align}
where $H\left( \omega \right)$ is a real-valued magnitude function, $L = 0$ for even symmetry of $\bf h$ and  $L = 1$ for odd symmetry of $\bf h$. Without loss of generality, in the following, we consider Type \uppercase\expandafter{\romannumeral1} filters ($N$ is odd and $L=0$), and hence $H\left( \omega \right)$ can be expressed as 
\begin{align}
    H\left( \omega \right) = &\sum_{n=0}^{\frac{N-3}{2}}2h\left[ n \right]\cos\left[ \left(\frac{N-1}{2} - n\right)\omega \right]\nonumber\\
    &+ h\left[ \frac{N-1}{2} \right]\label{eq: full_H}.
\end{align}

Furthermore, to obtain the optimal length $N$ frequency response with full-precision coefficients, one must solve the following minimax approximation problem:\cite{mitra2001digital}
\begin{align}
  \left(\mathcal{P}_2\right) \ E^*=   \underset{\mathbf{h}}{\min} \  \underset{\omega \in \Omega}{\max}\left( \left| W\left( \omega \right) \left[ H\left( \omega \right) -D\left( \omega \right) \right]  \right| \right), \label{eq: fir_e_optimal}
\end{align}
where $W\left( \omega \right)$ is the weighting function, $D\left( \omega \right)$ is the desired frequency response, and $\Omega$ is the set for passband and stopband intervals of the filter. The classic approach to solving problem $\left(\mathcal{P}_2\right)$ is the Parks–McClellan (PM) algorithm \cite{1083419}.

It is important to emphasize that solving problem $\left(\mathcal{P}_2\right)$ can only obtain full-precision optimal FIR filter coefficients. Nevertheless, using the FIR filter coefficients with full precision for hardware implementation is impractical, as finite wordlength effects must be taken into account \cite{1162497}.

Considering finite wordlength effects, the frequency response of the optimal FIR filter after quantization can be expressed as
\begin{align}
        \Hat{H}\left( e^{{j}\omega} \right) & = \sum_{n=0}^{N-1}{\Hat{h}\left[ n \right] e^{-{{j}}\omega n}}\\
        &= \sum_{n=0}^{N-1}{\mathcal{Q}\left(h\left[ n \right],b_n\right) e^{-{{j}}\omega n}} \\
    &= \Hat{H}\left( \omega \right) e^{{j}\left( \frac{L\pi}{2}-\frac{N-1}{2}\omega \right)},
\end{align}
where $\{\Hat{h}[n]=\mathcal{Q}\left(h\left[ n \right],b_n\right)\}_{n=0}^{N-1}$ are the optimal FIR filter coefficients after quantizing $h[n]$ using $b_n$ bits, $\Hat{H}\left( \omega \right)$ is the corresponding magnitude function, and $\mathcal{Q}\left(\cdot,b\right)$ is the $b$-bit fixed-point or floating-point rounding quantization function. Similar to \eqref{eq: full_H}, considering a Type \uppercase\expandafter{\romannumeral1} filter, $\Hat{H}\left( \omega \right)$ is given by 
\begin{align}
    \Hat{H}\left( \omega \right) = &\sum_{n=0}^{\frac{N-3}{2}}2\mathcal{Q}\left(h\left[ n \right],b_n\right)\cos\left[ \left(\frac{N-1}{2} - n\right)\omega \right]\nonumber\\
    &+ \mathcal{Q}\left(h\left[ \frac{N-1}{2} \right],b_{\frac{N-1}{2}}\right)\label{eq: finite_H}.
\end{align}
Additionally, the quantization bit sequence ${\bf b}=\{b_n\}_{n=0}^{N-1}$ is assumed to be even symmetry, i.e., $b_{n}=b_{N-1-n}, 0\leq n\leq N-1$, to preserve the linear-phase property of FIR filter after quantization.

Then, similar to problem $\left(\mathcal{P}_2\right)$, given the FIR filter coefficients and provided that linear-phase property after quantization, we can formulate the following mixed-precision minimax approximation problem to find the optimal bit allocation as
\begin{subequations}
\begin{align} 
{\left ({{{\mathcal{P}_3}} }\right)}~{\underset{\{ b_n \}_{n=0}^{N-1}}{\min}  }&~{\underset{\omega \in \Omega}{\max}\left( \left| W\left( \omega \right) \left[ \Hat{H}\left( \omega \right) -D\left( \omega \right) \right]  \right| \right)}  \label{eq: obj_p1}\\ 
{{\text {s.t.}}}~~\,&~{2\sum_{n=0}^{\frac{N-3}{2}}{b_n}+b_{\frac{N-1}{2}}\le N\cdot \bar{b},}  \label{eq: sum_budget}\\
&~\ b_n \in \mathbb{B}, \ \forall n = 0, 1, \dots, \frac{N-1}{2}.\label{eq: value_set}
\end{align}
\end{subequations}
Problem $\left(\mathcal{P}_3\right)$ is a challenging non-convex integer programming problem. We remark that it is the first time to consider the mixed-precision quantization for FIR filter design. Next, the proposed PSO-based algorithms will be utilized to solve problem $\left(\mathcal{P}_3\right)$.

\subsection{Proposed Algorithms}
Since problem $\left(\mathcal{P}_3\right)$ has the same format as the general mixed-precision quantization problem $\left(\mathcal{P}_1\right)$, we have 
\begin{align}
    F({\bf b}) &= \underset{\omega \in \Omega}{\max}\left( \left| W\left( \omega \right) \left[ \Hat{H}\left( \omega \right) -D\left( \omega \right) \right]  \right| \right), \label{eq: fir_fb}\\
    C({\bf b}) & = 2\sum_{n=0}^{\frac{N-3}{2}}{b_n}+b_{\frac{N-1}{2}}, \label{eq: fir_cb}\\
    C({\bar{b}}) & = N\cdot \bar{b}.\label{eq: fir_barb}
\end{align}
where ${\bf b}=\{ b_n \}_{n=0}^{N-1}$ and $\hat{H}\left( \omega \right)$ is given in \eqref{eq: finite_H}. Therefore, we can substitute \eqref{eq: fir_fb}, \eqref{eq: fir_cb} and \eqref{eq: fir_barb} into \textit{Algorithm} \ref{alg: PPSO} and \ref{alg: GC-PSO} to obtain the near-optimal bit allocation.

\begin{remark}[Complexity Analysis]\label{rem: cc_fir}
    The time complexity of the PPSO algorithm is $\mathcal{O}(N_{\rm pop}I_{\rm iter}N)$, which depends on the number of particles $N_{\rm pop}$, iteration number $I_{\rm iter}$ and the length of FIR filter $N$. Notably, the time complexity of \textit{Algorithm} \ref{alg: PPSO} increases linearly with $N$, much lower than that of brutal force search.
    Furthermore, the time complexity of the GC-PSO algorithm is $\mathcal{O}\left(N_{\rm pop}I_{\rm iter}rN^2\right)$, where $r$ is the total cycle number of \textit{Algorithm} \ref{alg: GC-PSO} in the GC-PSO algorithm. It is observed that the time complexity of the GC-PSO algorithm increases quadratically with $N$, which is more efficient than the brutal force search, but less than the PPSO algorithm. 
\end{remark}
\begin{remark}[Low-Complexity Solution]\label{rem: lc}
    Notably, the time complexity of the above two PSO-based algorithms is influenced by the number of particles and iteration number in addition to $N$. This complexity may remain significantly high when computing resources are severely limited. Consequently, we further propose low-complexity (LC) algorithms with time complexity of $\mathcal{O}\left(N\right)$ in Appendix \ref{app: LC}.
\end{remark}

\subsection{Numerical Example}
\label{sec: numerical}

\begin{table}[t]
\centering
\caption{Simulation Parameters for FIR Filter Design}
\label{tab: pso_para}
\begin{tabular}{@{}c|c|c|c@{}}
\toprule
\midrule
Parameters & Value & Parameters & Value \\ \midrule
   $I_{\rm iter}$        &   $100$    &       $\mathbb{B}$     &   $\{1,2,\cdots,2\Bar{b}+1\}$    \\
    $N_{\rm pop}$       &   $550$    &         $\lambda$   &  $10^3$     \\
    $[\omega_{\min},~\omega_{\max} ]$       &   $[0.4,~0.9]$    &      $[c_{1,\min},~c_{1,\max}]$      &   $[0.5,~2.5]$    \\
      $[v_{\min},~v_{\max} ]$     &  $[-3,~3]$     &     $[c_{2,\min},~c_{2,\max}]$       &   $[0.5,~2.5]$    \\ \midrule
\bottomrule
\end{tabular}
\end{table}

\begin{table}[t]
\centering
\caption{Filter Specifications}
\label{tab: filter}
\begin{tabular}{@{}c|c|c|c@{}}
\toprule
\midrule
Filter & Bands                                                                   & $D(\omega)$                                   & $W(\omega)$                                   \\ \midrule
A      & \begin{tabular}[c]{@{}c@{}}$[0,~0.4\pi]$\\ $[0.5\pi,~\pi]$\end{tabular} & \begin{tabular}[c]{@{}c@{}}1\\ 0\end{tabular} & \begin{tabular}[c]{@{}c@{}}1\\ 1\end{tabular} \\ \midrule
B      &  \begin{tabular}[c]{@{}c@{}}$[0,~0.4\pi]$\\ $[0.5\pi,~\pi]$\end{tabular}                                                                       &            \begin{tabular}[c]{@{}c@{}}1\\ 0\end{tabular}                                   &              \begin{tabular}[c]{@{}c@{}}1\\ 10\end{tabular}                                  \\ \midrule
C      &   \begin{tabular}[c]{@{}c@{}}$[0,~0.24\pi]$\\ $[0.4\pi,~0.68\pi]$\\ $[0.84\pi,~\pi]$\end{tabular}                                                                        &           \begin{tabular}[c]{@{}c@{}}1\\ 0\\ 1\end{tabular}                                    &                 \begin{tabular}[c]{@{}c@{}}1\\ 1\\1\end{tabular}                  \\ \midrule
D      &      \begin{tabular}[c]{@{}c@{}}$[0.02\pi,~0.42\pi]$\\ $[0.52\pi,~0.98\pi]$\end{tabular}                                           &                \begin{tabular}[c]{@{}c@{}}1\\ 0\end{tabular}                               &         \begin{tabular}[c]{@{}c@{}}1\\ 1\end{tabular}                       \\ \midrule
\bottomrule
\end{tabular}
\end{table}

We now present numerical results to demonstrate the superiority of the proposed algorithms. The MATLAB function $\mathtt{quantizer.m}$ and $\mathtt{quantize.m}$ is utilized to simulate fixed-point and floating-point quantization. The simulation parameters of the proposed algorithms are provided in Table \ref{tab: pso_para}. Moreover, we run \textit{Algorithm} \ref{alg: PPSO} and \ref{alg: GC-PSO} 10 times and choose the best results. The Type I FIR filter specifications are provided in Table \ref{tab: filter}, which is a classic setting in \cite{kodek2005telescoping,8307077}. Further, we use a combination of filter specification letter, filter length, and quantization bit to describe each filter in Table \ref{tab: filter}. For instance, A35/8 denotes a filter design with specification A using a length of $N=35$, and 8-bit fixed-point quantization ($\bar{b}=8$), while A35/[5,~4] indicates a 9-bit floating-point quantization with 5 bits for the exponent and 4 bits for the mantissa ($\bar{m}=4$). Based on these specifications, the full-precision FIR filter coefficients can be obtained using the MATLAB function $\mathtt{firpm.m}$.

 \begin{figure}[t]
    \centering
    \subfloat[Convergence curve of the proposed PPSO and GC-PSO algorithms with C35/8.]{\includegraphics[width=0.23\textwidth]{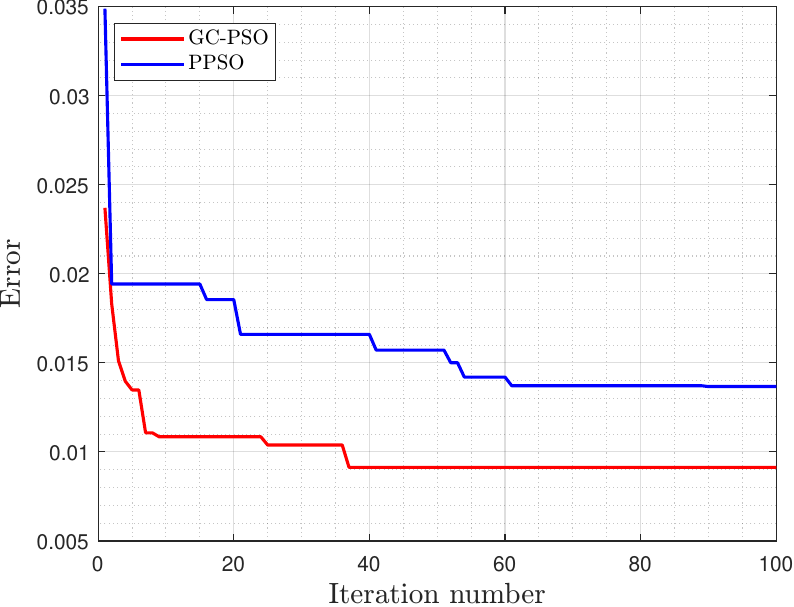}
    \label{fig: con_pso_fx}}
    \hfill
    \subfloat[The magnitude response of the filter A35/8 after fixed-point quantization.]{\includegraphics[width=0.22\textwidth]{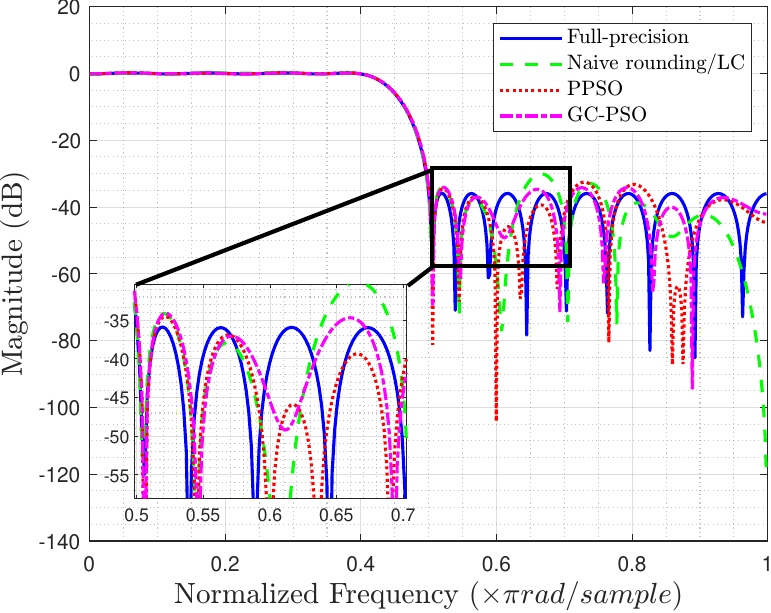}
    \label{fig: mag_fx}}
    \caption{Case study for fixed-point quantization.}
    \label{fig: fixed_p_fir}
\end{figure}


 
\begin{table*}[t]
\centering 
\setlength{\tabcolsep}{3.5pt}
\caption{Fixed-Point quantization error comparison for the filter specifications in Table \ref{tab: filter}}
\label{tab: fx}
\begin{tabular}{ccccccccc}
\toprule[1pt]
\midrule
 Filter & A35/8 & A45/8 & B35/9  & B45/9 & C35/8 & C45/8   & D35/8 & D45/8\\ \midrule
{Full-precision}   &   0.01595  & $7.132\cdot 10^{-3}$ & 0.05275 & 0.02111 & $2.631\cdot 10^{-3}$ & $6.709\cdot 10^{-4}$  & 0.01761 & $6.543\cdot10^{-3}$\\ \midrule
{Naive rounding/LC}    &  0.03266    & 0.03706 & 0.15879 & 0.11719 & 0.04687 & 0.03046  & 0.04692 & 0.03571 \\ \midrule
{Telescoping \cite{kodek2005telescoping}}   &  0.03266    & 0.03186 & 0.07854 & 0.06641 & 0.01787 & 0.02103   & 0.03404 & 0.03403\\ 
{LLL reduction \cite{8307077}}   &  0.02983 & 0.02962 & 0.08205 & 0.06041 & 0.01787 & 0.01609   & 0.03349 & 0.03167\\ 
{BKZ reduction \cite{8307077}}   &  0.02983 & 0.02962 & 0.08205 & 0.06041 & 0.01917 & 0.01609   & 0.03349 & 0.03094\\  
{HKZ reduction \cite{8307077}}   &  0.02983 & 0.02962 & 0.08205 & 0.06041 & 0.01917 & 0.02291   & 0.03349 & 0.02887\\ \midrule
\rowcolor{lightblue} {PPSO} & \underline{0.02364} & \underline{0.01450} & \underline{0.07677} & \underline{0.05948} & \underline{0.01367}& \underline{0.00751}  & \underline{0.02505} & \underline{0.01280}\\ \midrule
\rowcolor{cyan!20} {GC-PSO} & {\bf 0.02202} & {\bf 0.01182} & {\bf 0.07032} & {\bf 0.05058} & {\bf 0.00913}& {\bf 0.00554}  &{\bf 0.02194} & {\bf 0.01261}\\ \midrule
\bottomrule[1pt]
\end{tabular}
\end{table*}

\subsubsection{Fixed-Point Quantization}
In this subsection, we evaluate the performance of the proposed algorithms for fixed-point quantization. In Fig. \ref{fig: con_pso_fx}, we first analyze the convergence performance of the proposed PPSO and GC-PSO algorithms using C35/8 as an example. Both of them can converge to fixed values. Moreover, similar to the analysis in Section \ref{sec: ppso} and \ref{sec: gc-pso}, it can be observed that the GC-PSO algorithm can approach a lower error than the PPSO algorithm. As shown in Fig. \ref{fig: mag_fx}, naive rounding and the LC algorithm (See Appendix \ref{app: LC}) have the worst performance with maximum stopband attenuation of $-30.0983$ dB, higher than that of the PPSO algorithm ($-32.5597$ dB) and the GC-PSO algorithm ($-34.0964$ dB). These phenomena show that the proposed PPSO and GC-PSO algorithms can achieve better performance than that of naive rounding.

Then, to highlight the superiority of the proposed algorithms, we compare them with the most efficient quasi-optimal methods for fixed-point quantization, i.e., the telescoping rounding approach \cite{kodek2005telescoping} and the Lattice basis reduction approach \cite{8307077}, for different filter specifications. The results are shown in Table \ref{tab: fx}, where
\begin{itemize}
    \item the first row represents different filter specifications; 
    \item the second row lists the minimax error $E^*$ in \eqref{eq: fir_e_optimal} computed based on the PM algorithm with full-precision filter coefficients;
    \item the third row provides the errors in \eqref{eq: obj_p1} computed by direct rounding/LC algorithm of the filter coefficients;
    \item the fourth row gives the errors in \eqref{eq: obj_p1} obtained by using the telescoping rounding approach \cite{kodek2005telescoping};
    \item the fifth to seventh row shows the lattice-based quantization errors in \eqref{eq: obj_p1} when choosing the LLL, BKZ, and HKZ basis reduction option, respectively \cite{8307077};
    \item the last two rows gives the errors in \eqref{eq: obj_p1} from the proposed PPSO and GC-PSO algorithms.
\end{itemize}
In Table \ref{tab: fx}, bold values and underlined values denote the best and second-best results among all the algorithms except for the results from the full-precision filter coefficients, respectively. It is evident that the proposed PPSO and GC-PSO algorithms can achieve the optimal results for different filter specifications.

\begin{table*}[t]
\centering 
\setlength{\tabcolsep}{3.5pt}
\caption{Floating-Point quantization error comparison for the filter specifications given in Table \ref{tab: filter}}
\label{tab: fp}
\begin{tabular}{ccccccccc}
\toprule[1pt]
\midrule
 Filter & A35/[5,~4] & A45/[5,~4] & B35/[5,~5]  & B45/[5,~5] & C35/[5,~4] & C45/[5,~4]   & D35/[5,~4] & D45/[5,~4]\\ \midrule
{Full-precision}   &   0.01607  & $7.132\cdot 10^{-3}$ & 0.05312 & 0.02111 & $2.631\cdot 10^{-3}$ & $6.796\cdot 10^{-4}$  & 0.01761 & $6.543\cdot10^{-3}$\\ \midrule
{Naive rounding}    &  0.03738   & 0.03084 & 0.14556 & 0.11164 & 0.03955 & 0.03690 &  0.03500 & 0.02198  \\ \midrule
\rowcolor{SkyBlue!20} {LC}  &0.02341  & 0.01117 & 0.07884 & 0.03577 &0.01074 &0.00561  & 0.02121 & 0.01106 \\ \midrule
\rowcolor{lightblue} {PPSO} & \underline{0.01732} & \underline{0.00859} & \underline{0.05818} & \underline{0.02661} & \underline{0.00623}& \underline{0.00199} & \underline{0.02041} & \underline{0.00934} \\ \midrule
\rowcolor{cyan!20} {GC-PSO} & {\bf 0.01699} & {\bf 0.00842} & {\bf 0.05766} & {\bf 0.02520} & {\bf 0.00558}& {\bf 0.00197} & {\bf 0.02002} &{\bf 0.00817} \\ \midrule
\bottomrule[1pt]
\end{tabular}
\end{table*}

\begin{figure}[t]
    \centering
    \subfloat[Convergence curve of the proposed PPSO and GC-PSO algorithms with C35/[5,~4\text{]}]{\includegraphics[width=0.23\textwidth]{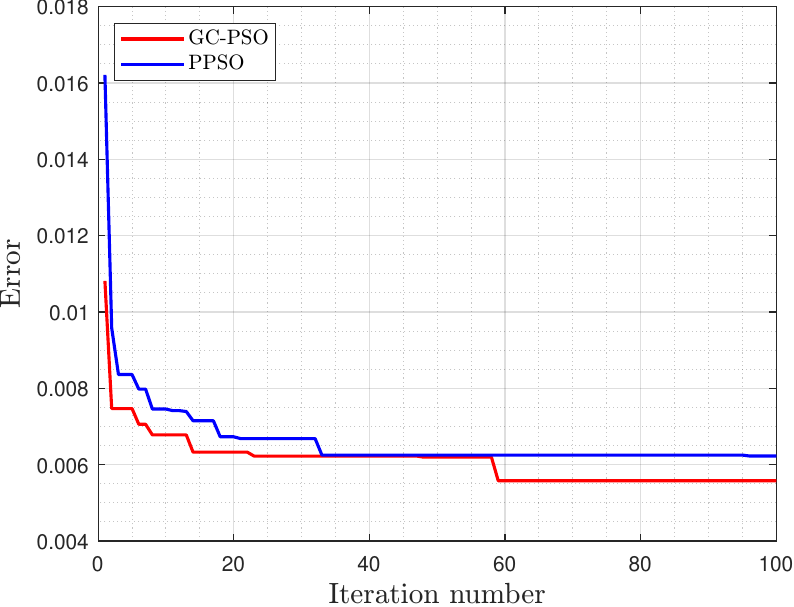}
    \label{fig: con_pso_fp}}
    \hfill
    \subfloat[The magnitude response of the filter A35/[5,~4\text{]} after fixed-point quantization.]{\includegraphics[width=0.22\textwidth]{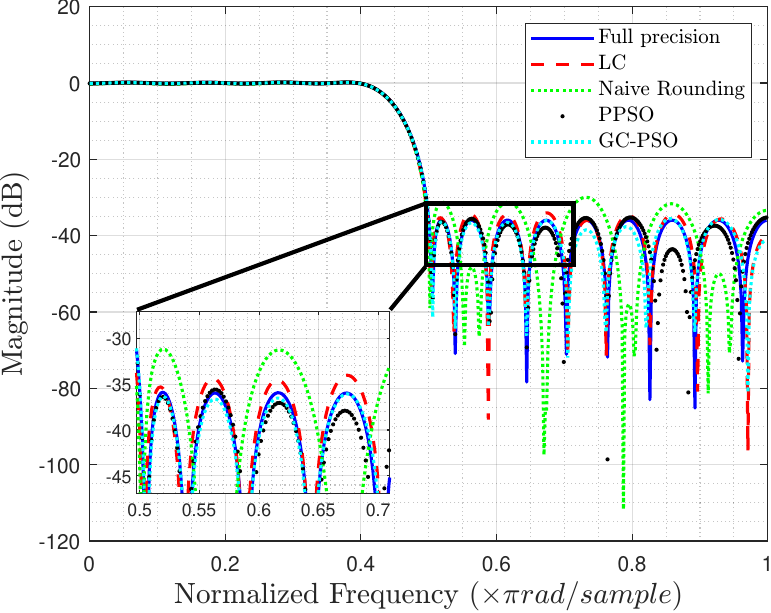}
    \label{fig: mag_fp}}
    \caption{Case study for floating-point quantization.}
    \label{fig: float_p_fir}
\end{figure}

\begin{table}[t]
\centering
\caption{Runtime of different algorithms with A45/[5,~4]}
\label{tab: runtime}
\begin{tabular}{@{}c|c|c|c@{}}
\toprule
\midrule
Algorithm & LC & PPSO & GC-PSO \\ \midrule
   Runtime (s)        &   $0.004$    &       $222.442$     &   $824.091$    \\
\midrule
\bottomrule
\end{tabular}
\end{table}



\subsubsection{Floating-Point Quantization}
In this subsection, we show the performance of the proposed algorithms for floating-point quantization. First, we examine the convergence performance of the proposed PPSO and GC-PSO algorithms using C35/[5,~4] as an example in Fig. \ref{fig: con_pso_fp}. It is shown that the GC-PSO algorithms can avoid the local optima but the PPSO algorithm can't. Moreover,  Fig. \ref{fig: mag_fp} presents the magnitude response of the filter A35/[5, 4] after floating-point quantization. Specifically, the maximum stopband attenuation of the LC algorithm, naive rounding, the PPSO algorithm, and the GC-PSO algorithm is $-34.0241$ dB, $-29.9598$ dB, $-35.2276$ dB, and $-35.5573$ dB, respectively. Therefore, the LC algorithm proposed in Section \ref{sec: so_fp} can achieve close optimal performance to the PPSO and GC-PSO algorithms and significantly better than that of naive rounding.

Then, we compare naive rounding, the LC algorithm, the PPSO algorithm, and the GC-PSO algorithm under various filter specifications, as shown in Table \ref{tab: fp}. The meaning of each row is similar to that of Table \ref{tab: fx}. The results reveal that the GC-PSO algorithm has the best performance. This is because the GC-PSO algorithm enables a better search of the feasible solution space. Additionally, although the LC algorithm's performance is inferior to that of the PPSO and GC-PSO algorithms, it has the lowest computational complexity (see Appendix \ref{sec: so_fp}), which can be advantageous when computational resources are limited. 

Finally, we compare the runtime of different algorithms using the A45/[5,~4] configuration as an example. All simulations are performed in MATLAB on a Windows machine equipped with a 13th Gen Intel Core i7-13700K processor (16 cores, 24 threads, 3.40 GHz) and 32 GB of DDR5 RAM. The runtimes of the various methods are reported in Table \ref{tab: runtime}. Consistent with the discussion in \textit{Remarks} \ref{rem: cc_fir} and \ref{rem: lc}, the GC-PSO algorithm exhibits the highest computational cost, while the LC algorithm achieves the shortest runtime.

\section{Mixed-Precision Application \uppercase\expandafter{\romannumeral2}: Receiver}
\label{sec: adc}
In this section, the proposed algorithms are applied to receivers in massive MIMO systems with precision-adaptive ADC architecture. First, we introduce the sum achievable rate maximum problem with precision-adaptive ADC. Then, the proposed algorithms are utilized to address it. Finally, the numerical example validates the performance of the proposed algorithms.

\subsection{Problem Statement}
\begin{figure}[t]
     \centering
     \includegraphics[width=0.3\textwidth]{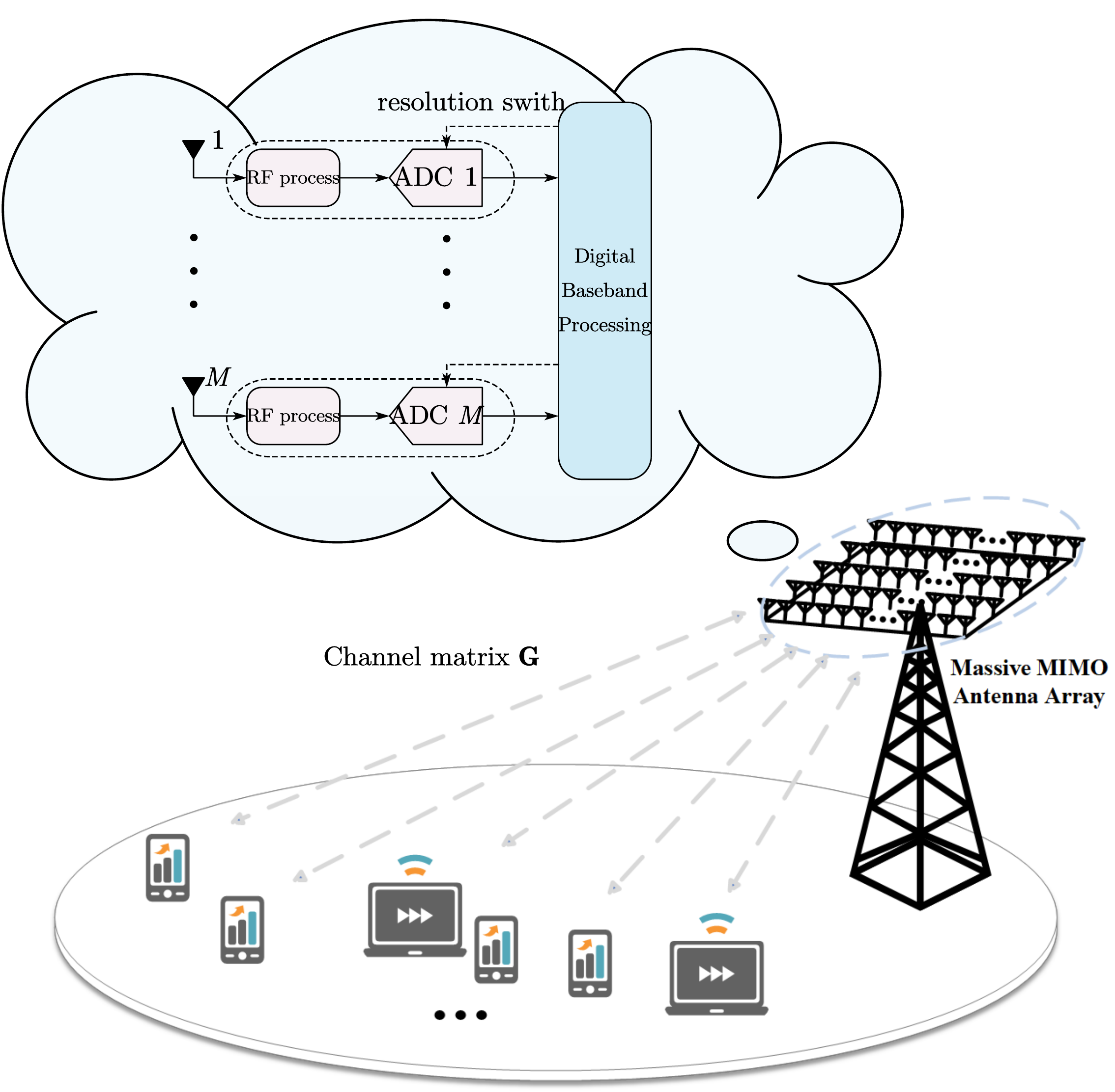}
     \caption{System model of uplink massive MIMO system with precision-adaptive ADC architecture.}
     \label{fig: adc}
\end{figure}
We consider an uplink single-cell massive MIMO system with $M$ antennas at the base station (BS) and $K$ single-antenna users as shown in Fig. \ref{fig: adc}. The received signal ${\bf y}\in \mathbb{C}^{M\times 1}$ is given by
\begin{align}
    {\bf y} = \sqrt{p_{\rm u}}{\bf G}{\bf x} + {\bf n},
\end{align}
where ${\bf x}\sim \mathcal{CN}(0,{\bf I}_K)$ is the transmitted signals from all the users, ${\bf n}\sim \mathcal{CN}(0,{\bf I}_M)$ is the additive white Gaussian noise (AWGN), and ${p_{\rm u}}$ is the average transmitted power of each user. ${\bf G}\in \mathbb{C}^{M\times N}$ is the channel matrix. We denote the channel coefficient between the $i$-th user and the $j$-th antenna of the BS as $g_{ji}=\sqrt{\gamma_i}h_{ji}$, where $h_{ji}\sim \mathcal{CN}(0,1)$ is the fast fading entry and $\gamma_i$ is the large-scale fading
coefficient \cite{6457363}. Further, in matrix form, we obtain
\begin{align}
    {\bf G} = {\bf HD}^{\frac{1}{2}},
\end{align}
where ${\bf H}\in \mathbb{C}^{M\times N}$ is the fast fading channel matrix and ${\bf D}={\rm diag}\left(\gamma_1,\gamma_2,\cdots,\gamma_K\right)$.

To alleviate the power consumption at the BS, a precision-adaptive ADC architecture is employed at the BS \cite{8017448,7886292}. Given that each of the $i$-th ADC pair has $b_i$ quantization bits and using the additive quantization noise model (AQNM) \cite{7307134}, the received signal after quantization can be expressed as
\begin{align}
    {\bf y}_{\rm q}&=\mathbf{Q}({\bf y})={\bf D}_\alpha{\bf y}+{\bf n}_{\rm q}\\
    &=\sqrt{p_{\rm u}}{\bf D}_\alpha{\bf G}{\bf x} + {\bf D}_\alpha{\bf n} +{\bf n}_{\rm q},
\end{align}
where $\mathbf{Q}(\cdot)$ is an element-wise ADC quantization function separately applied to the real and imaginary parts, ${\bf n}_{\rm q}$ is the quantization noise, and ${\bf D}_\alpha = {\rm diag}\left(\alpha_1,\alpha_2,\cdots,\alpha_M\right)$, $\alpha_i = 1-\beta_i$ is the quantization gain, where $\beta_i$ is a normalized quantization error satisfying Table \ref{tab: adc}. Moreover, for a fixed channel $\bf G$, the covariance matrix of quantization noise ${\bf n}_{\rm q}$ is 
\begin{align}
    {\bf R}_{{\bf n}_{\rm q}} = {\bf D}_\alpha{\bf D}_\beta{\rm diag}\left(p_{\rm u}{\bf G}{\bf G}^H + {\bf I}_{M}\right),
\end{align}
where ${\bf D}_\beta = {\rm diag}\left(\beta_1,\beta_2,\cdots,\beta_M\right)$.

\begin{table}[t]
\centering 
\setlength{\tabcolsep}{3.5pt}
\caption{The values of $\beta$ for different ADC quantization bits $b$}
\label{tab: adc}
\begin{tabular}{ccccccc}
\toprule[1pt]
\midrule
$b$ & 1 & 2 & 3 & 4 & 5 & $\geq 6$ \\ \midrule
$\beta$ & 0.3634 & 0.1175 & 0.03454 & 0.009497 & 0.002499 & $\frac{\pi \sqrt{3}}{2}2^{-2b_i}$\\ \midrule
\bottomrule[1pt]
\end{tabular}
\end{table} 

Furthermore, by applying the MRC receiver, the detected signal vector is given by
\begin{align}
    {\bf r} &= {\bf G}^H{\bf y}_{\rm q} \\
    & = \sqrt{p_{\rm u}}{\bf G}^H{\bf D}_\alpha{\bf G}{\bf x} + {\bf G}^H{\bf D}_\alpha{\bf n} +{\bf G}^H{\bf n}_{\rm q}. \label{eq: mrc}
\end{align}
Using \eqref{eq: mrc}, the received signal for the $k$-th user after detecting at the BS can be expressed as 
\begin{align}
    r_k &= \sqrt{p_{\rm u}} {\bf g}_k^H{\bf D}_\alpha{\bf g}_kx_k + \sqrt{p_{\rm u}}\sum_{i\neq k}^K {\bf g}_k^H{\bf D}_\alpha{\bf g}_ix_i \nonumber\\
    &+ {\bf g}_k^H{\bf D}_\alpha {\bf n} + {\bf g}_k^H{\bf n}_{\rm q}, \label{eq: mrc_n}
\end{align}
where ${\bf g}_k$ is the $k$-th column of the channel matrix $\bf G$. For a fixed $\bf G$, the last three terms in \eqref{eq: mrc_n} is the interference-plus-noise. Assuming the interference-plus-noise follows Gaussian distribution \cite{7886292}, we can obtain the ergodic achievable rate of the $k$-th user as follows:
\begin{align}
    R_k\left( \mathbf{b} \right) =\mathbb{E} \left[ \log _2\left( 1+\frac{p_u\left| \mathbf{g}_{k}^{H}\mathbf{D}_{\alpha}\mathbf{g}_k \right|^2}{\Phi} \right) \right], \label{eq: capacity_user}
\end{align}
where 
\begin{align}
    \Phi = p_u\sum_{i\ne k}^N{\left| \mathbf{g}_{k}^{H}\mathbf{D}_{\alpha}\mathbf{g}_i \right|^2}+\mathbf{g}_{k}^{H}\left( \mathbf{D}_{\alpha}^{2}+\mathbf{R}_{\mathbf{n}_q\mathbf{n}_q} \right) \mathbf{g}_k.
\end{align}

Then, based on \eqref{eq: capacity_user}, we can formulate the following sum achievable rate maximum problem with total ADC power consumption constraint. Specifically, we have
\begin{subequations}
\begin{align} 
{\left ({{{\mathcal{P}_4}} }\right)}~{\underset{\{ b_i \}_{i=1}^{M}}{\min}  }&~{-\sum_{k=1}^KR_k\left( \mathbf{b} \right)}  \label{eq: obj_p5}\\ 
{{\text {s.t.}}}~~\,&~{\sum_{i=1}^M{P_{\mathrm{ADC}}\left( b_i \right) \le M P_{\mathrm{ADC}}\left( \bar{b} \right)} ,}  \label{eq: adc_con}\\
&~\ b_i \in \mathbb{B}, \  i =  1,2, \cdots, M,\label{eq: adc_set}
\end{align}
\end{subequations}
where $P_{\mathrm{ADC}}(b_i) = cf_s2^{b_i}$, $c$ is the Walden’s figure-of-merit, and $f_s$ is the sampling rate \cite{761034}. Problem $\left ({{{\mathcal{P}_4}} }\right)$ is difficult to solve because it is implicit and non-convex with integer constraint. Notably, it has same format to the general mixed-precision quantization problem $\left ({{{\mathcal{P}_1}} }\right)$. Hence, in the subsequent subsection, we can apply the algorithms proposed in Section \ref{sec: mixed_precision_quantization} to address it.

\subsection{Proposed Algorithms}
Compared problem $\left ({{{\mathcal{P}_4}} }\right)$ with problem $\left ({{{\mathcal{P}_1}}}\right)$, we can obtain
\begin{align}
    F({\bf b}) &= -\sum_{k=1}^KR_k\left( \mathbf{b} \right), \label{eq: adc_fb}\\
    C({\bf b}) & = \sum_{i=1}^M P_{\mathrm{ADC}}\left( b_i \right), \label{eq: adc_cb}\\
    C({\bar{b}}) & = M P_{\mathrm{ADC}}\left( \bar{b} \right),\label{eq: adc_barb}
\end{align}
where ${\bf b}=\{ b_i \}_{i=1}^{M}$. Similarly, substituting \eqref{eq: adc_fb}, \eqref{eq: adc_cb} and \eqref{eq: adc_barb} into \textit{Algorithm} \ref{alg: PPSO} and \ref{alg: GC-PSO}, we can get the near-optimal bit allocation of ADCs. 

\begin{remark}[Complexity Analysis]
    The time complexity of the PPSO algorithm is given by $\mathcal{O}\left(N_{\rm pop}I_{\rm iter}MK^2\right)$, which relies on the number of particles $N_{\rm pop}$, the number of iterations $I_{\rm iter}$, is the number of BS antennas $M$, and the number of users $K$. Importantly, the time complexity of \textit{Algorithm} \ref{alg: PPSO} grows linearly with $M$, which is significantly lower than the brute-force search complexity of $\mathcal{O}\left(MK^2(\# \mathbb{B})^M\right)$.
    Moreover, the time complexity of the GC-PSO algorithm is $\mathcal{O}\left(N_{\rm pop}I_{\rm iter}rM^2K^2\right)$, where $r$ is the total cycle number of \textit{Algorithm} \ref{alg: GC-PSO} in the GC-PSO algorithm. It is observed that the time complexity of the GC-PSO algorithm increases quadratically with $M$, which is also more efficient than that of the brutal force search. 
\end{remark}

\subsection{Numerical Example}

\begin{table}[t]
\centering
\caption{Simulation Parameters for mixed ADC}
\label{tab: adc_para}
\begin{tabular}{@{}c|c|c|c@{}}
\toprule
\midrule
Parameters & Value & Parameters & Value \\ \midrule
   $I_{\rm iter}$        &   $100$    &       $\mathbb{B}$     &   $\{0,1,\cdots,2\Bar{b}+1\}$    \\
    $N_{\rm pop}$       &   $550$    &         $\lambda$   &  $10^3$     \\
    $[\omega_{\min},~\omega_{\max} ]$       &   $[0.4,~0.9]$    &      $[c_{1,\min},~c_{1,\max}]$      &   $[0.5,~2.5]$    \\
      $[v_{\min},~v_{\max} ]$     &  $[-3,~3]$     &     $[c_{2,\min},~c_{2,\max}]$       &   $[0.5,~2.5]$    \\ \midrule
\bottomrule
\end{tabular}
\end{table}

In this subsection, we present the simulation results to evaluate the performance of the proposed algorithms. The simulation parameters for the proposed algorithms are provided in Table \ref{tab: adc_para}. Note that we have $P_{\mathrm{ADC}}\left( b_i \right)  =0 $ and $\alpha_i = 0$ if $b_i =0$. In other words, the ADC pairs of the $i$-th antenna are deactivated. Moreover, we consider a scenario with $K=10$ users uniformly distributed within a hexagonal cell, where the BS is equipped with $M = 64$, and the cell radius is 1000 meters. The minimum distance between any user and the BS is set to $r_{\min} = 100$ meters \cite{7886292}. The path loss is modeled as $r_k^{-\nu}$, where $r_k$ is the distance between the $k$-th users and the BS, and $\nu = 3.8$ is the path loss exponent \cite{6816003}. Shadowing effects are represented by a log-normal random variable $o_k$ with a standard deviation $\sigma_o = 8$ dB. Therefore, the large-scale fading is given by $\gamma_k = o_k (r_k/r_{\min})^{-\nu}$. Moreover, for the HS algorithm in  \cite{10508306}, the number of initial solutions in the Harmony memory (HM) matrix is set to 550, matching the number of particles $N_{\rm pop}$ in GC-PSO. The Harmony memory considering rate (HMCR) is 0.9. The number of iterations is set to 30,000.

\begin{figure}[t]
    \centering
    \subfloat[Convergence curve of the proposed PPSO and GC-PSO algorithms with $\bar{b} = 1$, $M = 64$, $K=10$, and $p_{\rm u}=20~{\rm dB}$ for a fixed channel.]{\includegraphics[width=0.23\textwidth]{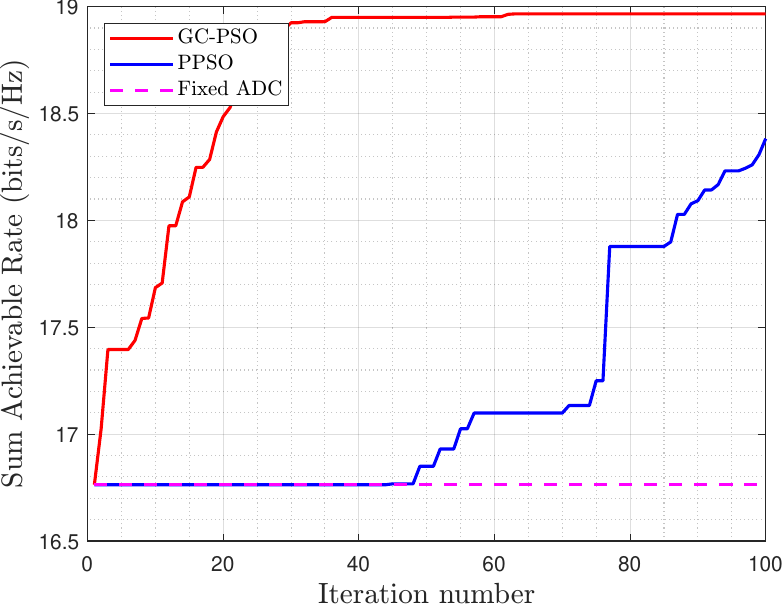}
    \label{fig: convergence_adc}}
    \hfill
    \subfloat[The sum achievable rate of precision-adaptive ADC with $\bar{b} = 1$, $M = 64$, $K=10$, and $p_{\rm u}=20~{\rm dB}$.]{\includegraphics[width=0.22\textwidth]{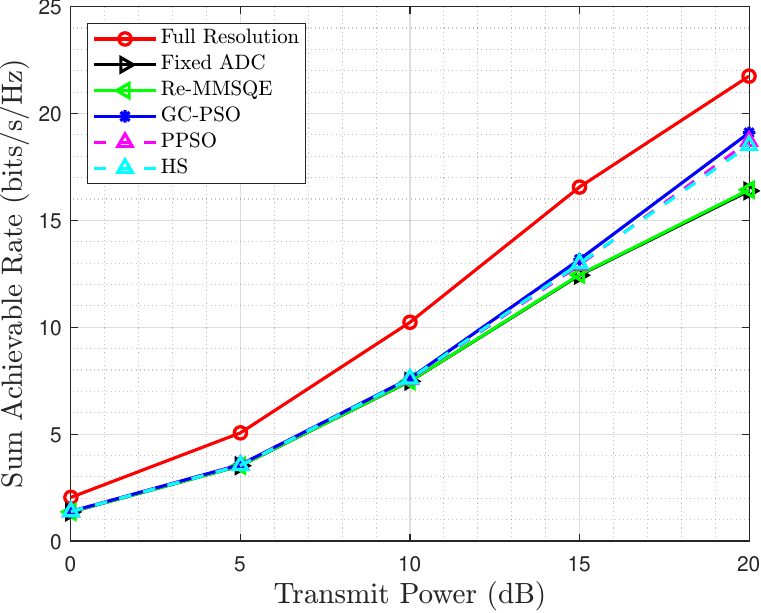}
    \label{fig: rate_adc}}
    \caption{Numerical example for the receiver in massive MIMO systems with precision-adaptive ADC architecture.}
    \label{fig: adc_simu}
\end{figure}



As shown in Fig. \ref{fig: convergence_adc}, we analyze the convergence performance of the proposed PPSO and GC-PSO algorithms with mixed-ADC under $\bar{b} = 1$, $M = 64$, $K=10$, and $p_{\rm u}=20~{\rm dB}$ for a fixed channel. It is observed that the GC-PSO algorithm requires fewer iterations to converge than the PPSO algorithm. Nevertheless, its complexity is higher due to the greedy criterion in each iteration. Additionally, the PPSO algorithm can achieve a better performance than the fixed-ADC system but with lower complexity.

To demonstrate the superiority of the proposed PPSO and GC-PSO algorithms, we compare them with the Re-MMSQE bit allocation method \cite{8017448} and the HS algorithm in \cite{10508306} using the sum achievable rate. Fig. \ref{fig: rate_adc} illustrates the sum achievable rate of the full-precision ADC, fixed-ADC, Re-MMSQE, GC-PSO, PPSO, and HS algorithms across different transmit powers with $\bar{b}=1$, $M=64$, $K=10$. It is evident that the Re-MMSQE, GC-PSO, and PPSO, and HS algorithms outperform the fixed-ADC system. Furthermore, the proposed PPSO and GC-PSO algorithms achieve $2$ dB gains over fixed-ADC at high transmit power levels, confirming their superiority.

\section{Mixed-Precision Application \uppercase\expandafter{\romannumeral3}: Gradient Descent}
\label{sec: GD}
As a final application, we use the proposed algorithms to address the quantization bit allocation for quantized GD. First, we introduce a minimum loss problem with a quantization resource budget at each iteration. Then, the proposed PPSO and GC-PSO algorithms are utilized to solve it. Finally, we present the simulations to validate the performance of the proposed algorithms.

\subsection{Problem Statement}
Consider the following optimization problem
\begin{align}
    \min_{\bf z}~f({\bf z}),\label{eq: gd_o}
\end{align}
where the objective loss function $f:\mathbb{R}^{{D}\times 1}\rightarrow \mathbb{R}$ is differentiable and ${\bf z} \in \mathbb{R}^{D\times 1}$ is the model parameter. 

  \begin{figure}[t]
     \centering
     \includegraphics[width=0.35\textwidth]{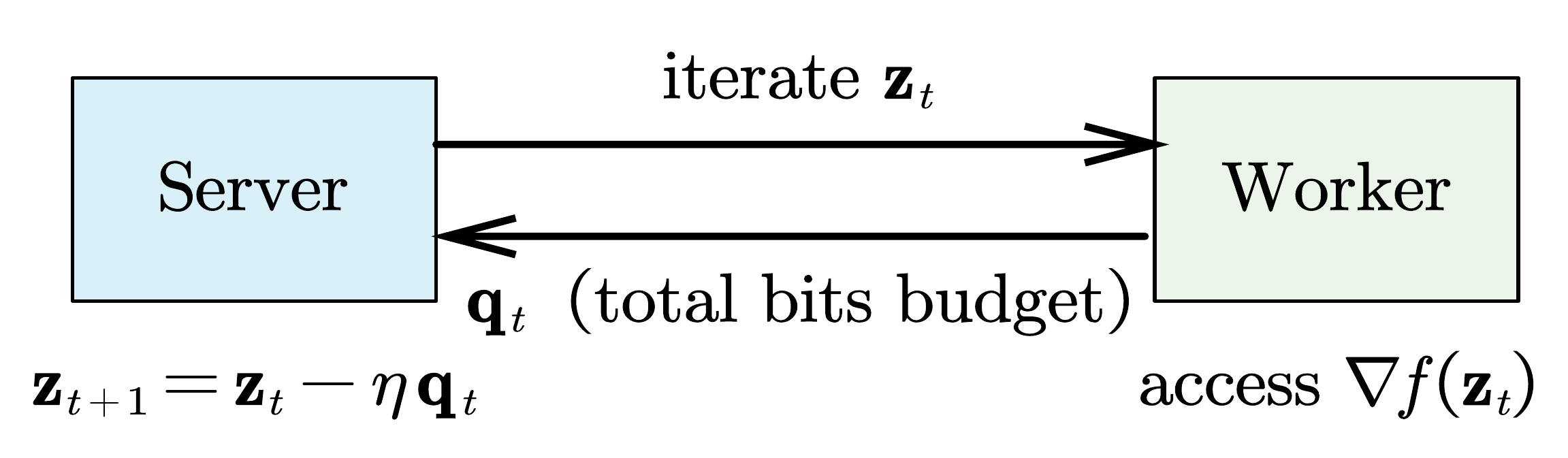}
     \caption{Quantized gradient descent (QGD) in a single-worker remote training setting.}
     \label{fig: qgd}
 \end{figure}
 
To find the optimal ${\bf z}^*$ in \eqref{eq: gd_o}, GD has been widely employed \cite{bekkerman2011scaling}. Following \cite{9764884}, we further consider a distributed scenario based on GD with a server and a worker as depicted in Fig. \ref{fig: qgd}. The server begins by transmitting the current iteration ${\bf z}_t$ to the worker without noise at the $t$-th iteration. The worker then computes the gradient $\nabla f\left({\bf z}_{t}\right)\in \mathbb{R}^{{D}\times 1}$. To reduce communication costs, a common approach is to quantize the gradient. In particular, for the $i$-th entry of the quantized gradient ${\bf q}_{t}$, we obtain
\begin{align}
    { q}_{t}^i = c_t\mathcal{Q}\left(\frac{[\nabla f\left({\bf z}_{t}\right)]_i}{c_t},b_i\right),\label{eq: gd_q}
\end{align}
where $\mathcal{Q}(\cdot)$ is the fixed-point quantization function by rounding to nearest, $[\nabla f\left({\bf z}_t\right)]_i$ is the $i$-th entry of the gradient, $b_i$ is the quantization bit for the $i$-th entry of the gradient and $c_t=\left\| \nabla f\left({\bf z}_{t}\right) \right\| _{2}$ is the normalized parameter. The worker then sends the quantized gradient ${\bf q}_t$ back to the server, which updates the model parameters according to the GD rule:
\begin{align}
    {\bf z}_{t+1} = {\bf z}_t - \eta {\bf q}_t,
\end{align}
where $\eta$ is the constant step size.

To find the optimal quantization bit allocation of the gradient, we can formulate the minimum loss function problem at each iteration $t+1$ under the total bits budget constraint as follows:
\begin{subequations}
\begin{align} 
{\left ({{{\mathcal{P}_5}} }\right)}~{\underset{\{ b_i \}_{i=1}^{D}}{\min}  }&~{f({\bf z}_t - \eta {\bf q}_t)}  \label{eq: obj_gd}\\ 
{{\text {s.t.}}}~~\,&~{{\sum_{i=1}^D b_i  \le D \bar{b} } ,}  \label{eq: gd_con}\\
&~\ b_i \in \mathbb{B}, \  i =  1,2, \cdots, D,\label{eq: gd_set}
\end{align}
\end{subequations}
where ${\bf q}_t$ is defined in \eqref{eq: gd_q}. Problem $\left ({{{\mathcal{P}_5}} }\right)$ is hard to be solved since \eqref{eq: obj_gd} is implicit and it involves in integer programming. Note that problem $\left ({{{\mathcal{P}_5}} }\right)$ is actually a particular example of the general mixed-precision quantization problem $\left ({{{\mathcal{P}_1}} }\right)$. Consequently, the proposed algorithms in Section \ref{sec: mixed_precision_quantization} can be utilized to address problem $\left ({{{\mathcal{P}_5}} }\right)$.

\subsection{Proposed Algorithms}
Comparing problem $\left (\mathcal{P}_5\right)$ with problem $\left (\mathcal{P}_1\right)$, we have the following expressions: 
\begin{align}
    F({\bf b}) &= f({\bf z}_t - \eta {\bf q}_t), \label{eq: gd_fb}\\
    C({\bf b}) & = \sum_{i=1}^D b_i, \label{eq: gd_cb}\\
    C({\bar{b}}) & = D \bar{b},\label{eq: gd_barb}
\end{align}
where ${\bf b}=\{ b_i \}_{i=1}^{D}$. Then, substituting \eqref{eq: gd_fb}, \eqref{eq: gd_cb} and \eqref{eq: gd_barb} into \textit{Algorithm} \ref{alg: PPSO} and \ref{alg: GC-PSO}, we can get the optimal bit allocation of quantized GD. 

\begin{remark}[Complexity Analysis]
    When solving problem $\left (\mathcal{P}_5\right)$ via brute-force search, the time complexity increases exponentially with the dimension of model parameters $D$. In contrast, the time complexity of the PPSO algorithm is $\mathcal{O}(N_{\rm pop}I_{\rm iter}D)$, which scales linearly with $D$ and is significantly lower than that of brute-force search. The time complexity of the GC-PSO algorithm is $\mathcal{O}\left(N_{\rm pop}I_{\rm iter}rD^2\right)$, where $r$ is the total cycle number of \textit{Algorithm} \ref{alg: GC-PSO} within the GC-PSO algorithm. Thus, the GC-PSO algorithm exhibits a quadratic complexity with respect to $D$, making it more efficient than brute-force search.
\end{remark}

\subsection{Numerical Example}
\begin{table}[t]
\centering
\caption{Simulation Parameters for GD}
\label{tab: gd_para}
\begin{tabular}{@{}c|c|c|c@{}}
\toprule
\midrule
Parameters & Value & Parameters & Value \\ \midrule
   $I_{\rm iter}$        &   $100$    &       $\mathbb{B}$     &   $\{1,2,\cdots,2\bar{b}+1\}$    \\
    $N_{\rm pop}$       &   $550$    &         $\lambda$   &  $10^5$     \\
    $[\omega_{\min},~\omega_{\max} ]$       &   $[0.4,~0.9]$    &      $[c_{1,\min},~c_{1,\max}]$      &   $[0.5,~2.5]$    \\
      $[v_{\min},~v_{\max} ]$     &  $[-3,~3]$     &     $[c_{2,\min},~c_{2,\max}]$       &   $[0.5,~2.5]$    \\ \midrule
\bottomrule
\end{tabular}
\end{table}

In this subsection, we evaluate the performance of the proposed algorithms through numerical examples based on the least squares
problem and logistic regression for binary classification. The simulation parameters of the proposed GC-PSO and PPSO algorithms are shown in Table \ref{tab: gd_para}.

\begin{figure}[t]
    \centering
    \subfloat[Convergence curve of the proposed algorithms and naive rounding with $\bar{b} = 4$, $\eta = 0.001$ and Gaussian matrix ($T=1000,D=100$).]{\includegraphics[width=0.23\textwidth]{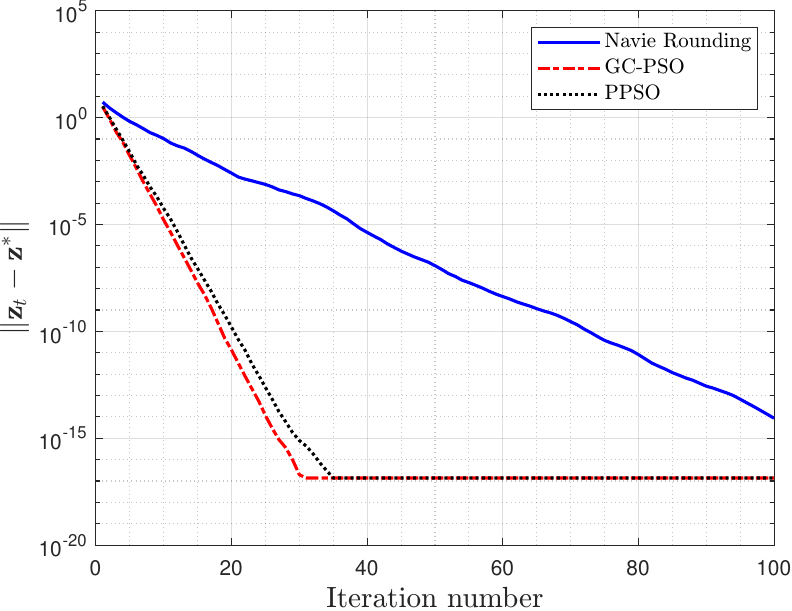}
    \label{fig: convergence_gd_g}}
    \hfill
    \subfloat[Convergence curve of the proposed algorithms and naive rounding with $\bar{b} = 4$, $\eta = 0.01$ and matrix $\mathtt{ash331}$ ($T=331,D=104$).]{\includegraphics[width=0.23\textwidth]{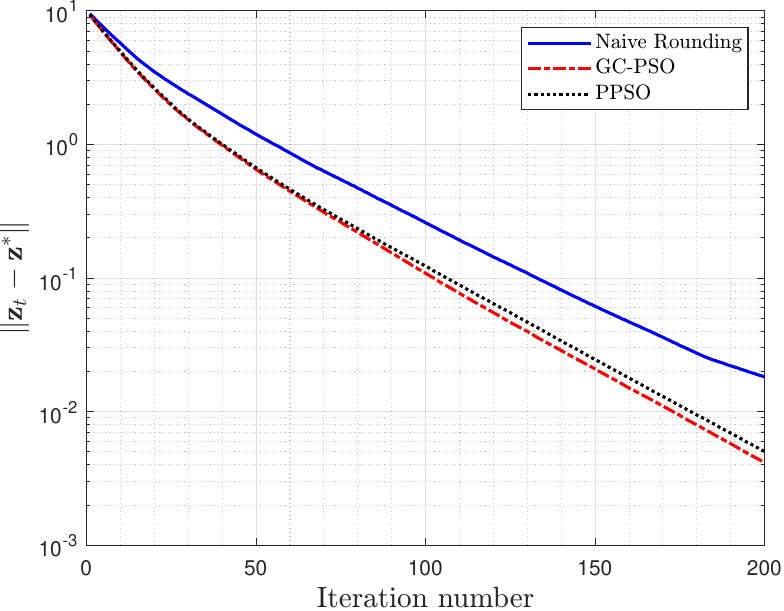}
    \label{fig: convergence_gd_a}}
    \caption{Numerical example for least squares problem.}
    \label{fig: gd_ls}
\end{figure}


 
\subsubsection{Least Squares} For the least squares problem, we have
\begin{align}
    f\left( \mathbf{z} \right) =\frac{1}{2}\left\| \mathbf{y}-\mathbf{Az} \right\| _{2}^{2},\notag
\end{align}
where ${\bf y}\in \mathbb{R}^{D\times 1}$ and ${\bf A} \in \mathbb{R}^{T\times D}$ with $T\geq D$. We first generate $\bf A$ with independently and identically distributed (i.i.d.) standard normal entries. Additionally, we use the real-world least squares matrix \texttt{ash331} as $\bf A$, obtained from the online repository SuiteSparse \cite{kolodziej2019suitesparse}. Then we sample ${\bf z}^*$ from $\mathcal{N}(0,1)$ and set $\mathbf{y} = \mathbf{Az}^*$. Moreover, the initial ${\bf z}_0$ is set to be a zero vector.

The convergence performance of the proposed PPSO and GC-PSO algorithms is measured by the error term $\left\|\mathbf{z}_t-\mathbf{z}^* \right\|_2$, where $\mathbf{z}_t$ is the computed parameter at the end of $t$-th iteration of quantized GD. As shown in Fig. \ref{fig: convergence_gd_g} and \ref{fig: convergence_gd_a}, the proposed PPSO and GC-PSO algorithms achieve faster convergence than naive rounding (fixed-precision quantization).


\begin{figure}[t]
    \centering
    \subfloat[$\bar{b}=2$.]{\includegraphics[width=0.23\textwidth]{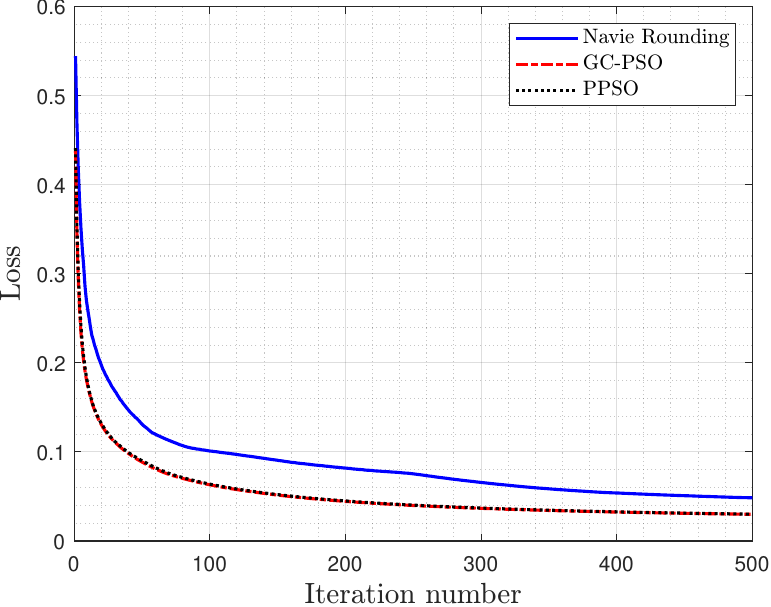}
    \label{fig: convergence_gd_bar2}}
    \hfill
    \subfloat[$\bar{b}=4$.]{\includegraphics[width=0.23\textwidth]{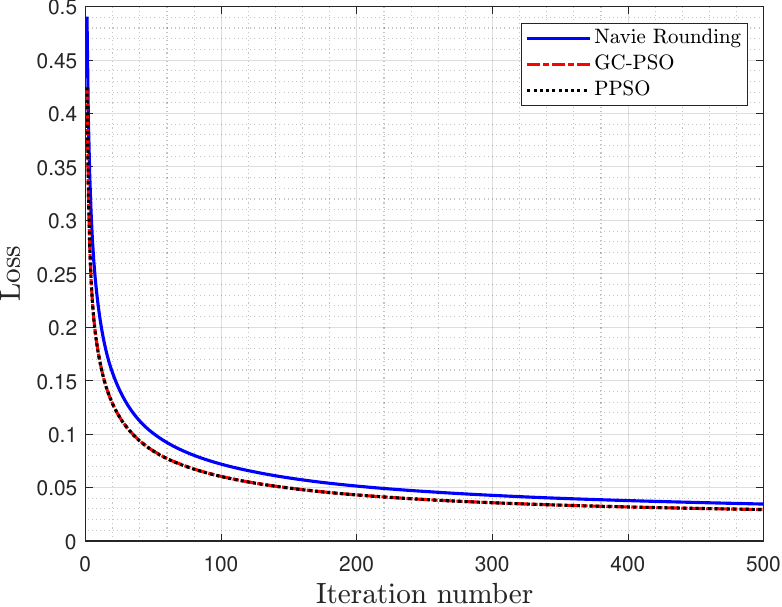}
    \label{fig: convergence_gd_bar4}}
    \caption{Training loss of the proposed algorithms and naive rounding with $\eta = 0.5$ and WBC diagnosis task.}
    \label{fig: convergence_gd_log}
\end{figure}
 
\subsubsection{Binary Classification} We further compare the proposed PPSO and GC-PSO algorithms with naive rounding for the binary classification problem with logistic regression. The logistic regression objective function is given by \cite{koh2007interior}
\begin{align}
    f\left( \mathbf{z} \right) = \frac{1}{m}\sum_{i=1}^{m}\log\left(1+\exp{\left(-y_i\cdot {\bf z}^T{\bf v}_i\right)} \right) + \frac{1}{2m}\left\| \mathbf{z} \right\| _{2}^{2},\notag
\end{align}
where $m$ is the size of train sets, ${\bf v}_i \in \mathbb{R}^{D\times 1}$ is the feature vector, and $y_i \in \{-1,1\}$ is the corresponding binary label.

The Wisconsin Breast Cancer (WBC) diagnosis task \cite{chang2011libsvm} is utilized as an example. We use the standard training and testing procedures \cite{lecun2002efficient}. For each $i$ sample from the dataset, the feature ${\bf v}_i$ dimension is $D = 30$, and each label $y_i$ is a binary number. The training set is repeatedly presented, with samples in random order. We set $m = 3000$ and $500$ samples to train and test, respectively. Fig. \ref{fig: convergence_gd_log} displays the training loss of the proposed algorithms compared to naive rounding, demonstrating the superiority of the proposed PPSO and GC-PSO algorithms with different total bit budgets. Moreover, the proposed PPSO and GC-PSO algorithms achieve $98.05\%$ accuracy, higher than $97.46\%$ accuracy of naive rounding on the test data.

\section{Conclusions}
\label{sec: con}
In this paper, we have proposed a bit allocation framework for mixed-precision quantization. First, we have formulated a general bit allocation problem for mixed-precision quantization. To address the integer consumption constraint, we have introduced the PPSO algorithm. Then, we have proposed a GC-PSO algorithm to avoid spending iterations on these infeasible solutions in the PPSO algorithm. Furthermore, the search framework has been applied to different fields, including FIR filter design, receivers, and GD. Finally, the search framework have achieved better performance compared with other algorithms in particular applications. For example, in the application of receivers, the proposed framework has achieved $2$ dB gains compared with fixed-ADC at high transmit power levels.

\appendices
\section{Proof of \textit{Theorem} \ref{the: con}}
\label{app: the}
    For simplification, we denote 
    \begin{align}
        &c=\frac{c_1+c_2}{2},~ b_{\mathrm{best}}=\frac{c_1}{c_1+c_2}b_{p}^{\mathrm{best}}+\frac{c_2}{c_1+c_2}b_{\mathrm{g}}^{\mathrm{best}},\\
        &x^{\mathrm{it}} = b^{\mathrm{it}} - b_{\mathrm{best}},
    \end{align}
    where $b_{\mathrm{best}}$ is set to be a constant. Then \eqref{eq: ve_one} and \eqref{eq: po_one} is simplified to
    \begin{align}
        v^{\mathrm{it}+1}&=wv^{\mathrm{it}}-cx^{\mathrm{it}}, \label{eq: ve_s}\\
        x^{\mathrm{it}+1}&=x^{\mathrm{it}}+\mathtt{round}\left( v^{\mathrm{it}+1} \right) 
         = x^{\mathrm{it}} +  v^{\mathrm{it}+1} + \epsilon^{\mathrm{it+1}}(v), \label{eq: po_s}
    \end{align}
    where $\epsilon^{\mathrm{it+1}}(v) = \mathtt{round}\left( v^{\mathrm{it}+1} \right) -v^{\mathrm{it}+1}$ is the rounding error satisfying $\left|\epsilon^{\mathrm{it+1}}(v)\right| \leq \left|v^{\mathrm{it}+1}\right|$. 

    Further, assuming a continuous process \cite{985692}, \eqref{eq: ve_s} and \eqref{eq: po_s} become differential equations as 
    \begin{align}
        \frac{\mathrm{d}v\left( t \right)}{\mathrm{d}t}&=\left( w-1 \right) v\left( t \right) -cx\left( t \right),\label{eq: ve_d}\\
        \frac{\mathrm{d}x\left( t \right)}{\mathrm{d}t}&=v\left( t+1 \right) + \epsilon \left( t+1,v \right), \label{eq: po_d}
    \end{align}
    where $\left|\epsilon \left( t+1,v \right)\right|\leq \left|v(t+1)\right|$. 

    Making a first-order approximation of $v(t+1)$, i.e., $v(t+1)=v(t)+\frac{\mathrm{d}v\left( t \right)}{\mathrm{d}t}$, and substituting it into \eqref{eq: po_d}, we obtain
    \begin{align}\label{eq: po_d_2}
        \frac{\mathrm{d}x\left( t \right)}{\mathrm{d}t}=wv\left( t \right) -cx\left( t \right) +\epsilon \left( t,v \right), 
    \end{align}
    where $\left|\epsilon \left( t,v \right)\right|\leq \left|wv(t)-cx(t)\right|$.

    Next, we combine \eqref{eq: ve_d} and \eqref{eq: po_d_2} and written in compact matrix form as follows:
    \begin{align}
        \dot{\mathbf{y}}=\mathbf{Ay}+\Delta \left( \mathbf{y} \right), \label{eq: pertubed}
    \end{align}
    where 
    \begin{align}
           & \mathbf{y}=\left[ \begin{array}{c}
	v\\
	x\\
\end{array} \right] , \mathbf{A}=\left[ \begin{matrix}
	w-1&		-c\\
	w&		-c\\
\end{matrix} \right],
\Delta \left( \mathbf{y} \right) =\left[ \begin{array}{c}
	0\\
	\epsilon \left( t,v \right)\\
\end{array} \right],
    \end{align}
and $ \dot{\mathbf{y}}$ is the derivative of $\mathbf{y}$. Note that \eqref{eq: pertubed} can be viewed as a dynamical perturbed system, where $\bf A$ is the state matrix in dynamical system theory, and $\Delta \left( \mathbf{y} \right)$ is perturbation term. ${\bf y=0}$ is an equilibrium point of the perturbed system. Therefore, we can transform the original convergence analysis into the stability analysis of a perturbed dynamical system. 

First, considering the nominal system, i.e., the dynamical system in \eqref{eq: pertubed} without perturbation term $\Delta \left( \mathbf{y} \right)$, the stability or convergence property depends on the eigenvalues of the state matrix $\bf A$. Specifically,
\begin{align}
    &\left| \lambda \mathbf{I}-\mathbf{A} \right|=0,\\
    \Longrightarrow &\lambda _{1,2}=\frac{\left( w-c-1 \right) \pm \sqrt{\left( w-c-1 \right) ^2-4c}}{2},
\end{align}
where $\lambda _{1,2}$ is the two eigenvalues of the state matrix $\bf A$. The necessary and sufficient condition for convergence, i.e., the equilibrium point ${\bf y=0}$ of the nominal system is stable, is that $\Re \left( \lambda _{1,2} \right) <0$ \cite[Theorem 4.5]{khalil2002nonlinear} and note that $w>0$ \cite{985692}, leading to the result
\begin{align}
	0<w<c+1.\label{eq: condition_1}
\end{align}


Second, for the dynamical perturbed system in \eqref{eq: pertubed}, a common approach to determining the stable condition is using the Lyapunov function $V\left( \mathbf{y} \right)$. Specifically, the original stable equilibrium point ${\bf y=0}$ of the nominal system is a stable equilibrium point of the perturbed system if the derivative of $V\left( \mathbf{y} \right)$ is negative \cite[Theorem 4.2 \& Lemma 9.1]{khalil2002nonlinear}. Therefore, we next determine the Lyapunov function $V\left( \mathbf{y} \right)$.

Since $\bf A$ is Hurwitz under the condition \eqref{eq: condition_1}, we can solve the following Lyapunov equation
\begin{align}
    \mathbf{PA}+\mathbf{A}^T\mathbf{P}=-\mathbf{I},~\mathbf{P}=\mathbf{P}^T,
\end{align}
and obtain a unique solution as
\begin{align}
    \mathbf{P}=\left[ \begin{matrix}
	\frac{c+c^2+w^2}{2c\left( 1+c-w \right)}&		\frac{-c^2+w-w^2}{2c\left( 1+c-w \right)}\\
	\frac{-c^2+w-w^2}{2c\left( 1+c-w \right)}&		\frac{1+c+c^2-2w+w^2}{2c\left( 1+c-w \right)}\\
\end{matrix} \right].
\end{align}
And the Lyapunov function is given by
\begin{align}
    V\left( \mathbf{y} \right) =\mathbf{y}^T\mathbf{Py}.
\end{align}

Then, the derivative of $V\left( \mathbf{y} \right)$ along the trajectories of the perturbed system satisfies
\begin{align*}
    \dot{V}\left( \mathbf{y} \right) &=\frac{\partial V}{\partial \mathbf{y}}\mathbf{Ay}+\frac{\partial V}{\partial \mathbf{y}}\Delta \left( \mathbf{y} \right) =2\mathbf{y}^T\mathbf{PAy}+2\mathbf{y}^T\mathbf{P}\Delta \left( \mathbf{y} \right) \\ 
    &\overset{\left( b \right)}{=}-\mathbf{y}^T\mathbf{y}+2\mathbf{y}^T\mathbf{P}\Delta \left( \mathbf{y} \right) \\
    &\le -\left\| \mathbf{y} \right\| _{2}^{2}+2\left\| \mathbf{P} \right\| _2\left\| \mathbf{y} \right\| _2\left\| \Delta \left( \mathbf{y} \right) \right\| _2\\
    &\overset{\left( c \right)}{\leq}-\left\| \mathbf{y} \right\| _{2}^{2}+2\lambda _{\max}\left( \mathbf{P} \right)\sqrt{c^2+w^2} \left\| \mathbf{y} \right\| _2^2\\
    &= \left(2\sqrt{c^2+w^2}\lambda _{\max}\left( \mathbf{P} \right)-1\right)\left\| \mathbf{y} \right\| _2^2<0,
\end{align*}
where we have $2\mathbf{y}^T\mathbf{PAy}=\mathbf{y}^T\mathbf{PAy}+\mathrm{tr}\left( \mathbf{y}^T\mathbf{PAy} \right) =\mathbf{y}^T\mathbf{PAy}+\mathrm{tr}\left( \mathbf{y}^T\mathbf{A}^T\mathbf{Py} \right) =\mathbf{y}^T\left( \mathbf{PA}+\mathbf{A}^T\mathbf{P} \right) \mathbf{y}=-\mathbf{y}^T\mathbf{y}$ at $(b)$, and $\left\| \Delta \left( \mathbf{y} \right) \right\| _2=\left| \epsilon \left( t,v \right) \right|\le\left| wv-cx \right| \le \sqrt{c^2+w^2} \left\|  \mathbf{y}  \right\| _2$ at $(c)$. Hence, the perturbed system is stable if 
\begin{align}
    \lambda _{\max}\left( \mathbf{P} \right) <\frac{1}{2\sqrt{c^2+w^2}} \label{eq: condition_2}.
\end{align}

In summary, based on condition \eqref{eq: condition_1} and \eqref{eq: condition_2}, the dynamical perturbed system in \eqref{eq: pertubed} is stable and converges to an equilibrium point. And the proof ends.
 
\section{Proposed Low-Complexity Solutions for Mixed-Precision FIR Filter Design}
\label{app: LC}
To reduce the time complexity of \eqref{eq: obj_p1}, by assuming that the quantization errors for the filter coefficients are independent random variables\footnote{It is a classic assumption for the analysis of the effect of coefficient quantization on filter response, even though for a given filter the quantization process is performed only once \cite{1162497,mitra2001digital}.}, we can transform problem $\left(\mathcal{P}_{3}\right)$ into the following MMSE problem:
\begin{align} 
{\left ({{{\mathcal{P}_6}} }\right)}~{\underset{\{ b_n \}_{n=0}^{N-1}}{\min}  }&~{\mathbb{E} \left\{ \int_0^{\pi}{\left| \hat{H}\left( \omega \right) -H\left( \omega \right) \right|^2d\omega} \right\}}  \label{eq: obj_p2}\\ 
{{\text {s.t.}}}~~\,&~ \eqref{eq: sum_budget},~\eqref{eq: value_set}, \nonumber
\end{align}
where ${H}\left( \omega \right)$ and $\hat{H}\left( \omega \right)$ is given in \eqref{eq: full_H} and \eqref{eq: finite_H}, respectively.

Since the quantization errors of fixed-point and floating-point quantization are different \cite{5995137}, we propose two solutions to address problem $\left(\mathcal{P}_{6}\right)$ for fixed-point and floating-point quantization, respectively. 

\subsubsection{Solution for Fixed-Point Quantization}
\label{sec: so_fx}
First, we give the following lemma for fixed-point quantization.
\begin{lemma}[Fixed-point quantization model \cite{1162497}]\label{lem: fixed_quant_error}
    For $b_n+1$ bit fixed-point quantization, given input filter coefficient $h[n]$, the output $\Hat{h}[n]$ is given by
    \begin{align}
        \Hat{h}[n] = \boldsymbol{q}(h[n],b_n) = h[n] + e_n,
    \end{align}
    where $\boldsymbol{q}(\cdot)$ is the fixed-point quantization function, and the quantization error $e_n$ satisfies uniform distribution with zero mean and ${2^{-2b_n}}/{12}$ variance.
\end{lemma}
Based on \textit{Lemma} \ref{lem: fixed_quant_error}, \eqref{eq: finite_H} can be expressed as
\begin{align}
    \Hat{H}\left( \omega \right) = &\sum_{n=0}^{\frac{N-3}{2}}2\left(h\left[ n \right]+e_n \right)\cos\left[ \left(\frac{N-1}{2} - n\right)\omega \right]\nonumber\\
    &+ \left(h\left[ \frac{N-1}{2} \right]+e_{\frac{N-1}{2}}\right) \label{eq: finite_H2}.
\end{align}
Using \eqref{eq: finite_H2} and considering the quantization errors for different filter coefficients are independent, the objective function \eqref{eq: obj_p2} can be simplified as follows: 
\begin{align}
    &\mathbb{E} \left\{ \int_0^{\pi}{\left| \hat{H}\left( \omega \right) -H\left( \omega \right) \right|^2d\omega} \right\} =\sum_{n=0}^{\frac{N-3}{2}}{\frac{\pi}{6}2^{-2b_n}}+\frac{\pi}{12}2^{-2b_{\frac{N-1}{2}}}.\label{eq: obj_p2.1}
\end{align}

Then we can transform problem $\left(\mathcal{P}_{6}\right)$ into
\begin{align*} 
{\left ({{{\mathcal{P}_{6.1}}} }\right)}~{\underset{\{ b_n \}_{n=0}^{N-1}}{\min}  }&~{\eqref{eq: obj_p2.1}} ~~~ 
{{\text {s.t.}}}~ \eqref{eq: sum_budget},~\eqref{eq: value_set}.
\end{align*}
To avoid integer programming, we relax the integer variables ${\bf b}\in \mathbb{B}^{N\times 1}$ to the real numbers $\Tilde{\bf b}\in \mathbb{R}^{N\times 1}$ to find a closed-form solution. Specifically, the relaxed problem can be expressed as 
\begin{align*} 
{\left ({{{\mathcal{P}_{6.2}}} }\right)}~{\underset{\{ \Tilde{b}_n \}_{n=0}^{N-1}}{\min}  }&~{\sum_{n=0}^{\frac{N-3}{2}}{\frac{\pi}{6}2^{-2\Tilde{b}_n}}+\frac{\pi}{12}2^{-2\Tilde{b}_{\frac{N-1}{2}}}} \\
{{\text {s.t.}}}~~\,&~{2\sum_{n=0}^{\frac{N-3}{2}}{\Tilde{b}_n}+\Tilde{b}_{\frac{N-1}{2}}\le N\cdot \bar{b}.}
\end{align*}

Furthermore, the following proposition provides a closed-form solution by solving the Karush-Kuhn-Tucker (KKT) conditions \cite{boyd2004convex} for problem $\left(\mathcal{P}_{6.2}\right)$. 
\begin{proposition}[Closed-form solution for the relaxed fixed-point quantization problem]\label{prop: fixed}
    For problem $\left(\mathcal{P}_{6.2}\right)$, the optimal bit allocation is derived as 
    \begin{align}\label{eq: prop_1}
        \Tilde{ b}_n = \Bar{b},~n = 0,1,\cdots,\frac{N-1}{2}.
    \end{align}
\end{proposition}
\begin{IEEEproof}
    By denoting $z_n=2^{2\Tilde{b}_n},~n=0,1,\cdots,\frac{N-1}{2}$, $\Bar{z}=2^{-2\Bar{b}}$, $c_n = \frac{\pi}{6},~n=0,1,\cdots,\frac{N-3}{2}$ and $c_{\frac{N-1}{2}}=\frac{\pi}{12}$, we can convert the problem $\left(\mathcal{P}_{6.2}\right)$ into a simpler form given by
    \begin{subequations}\label{eq: sim_fixed}
            \begin{align}
        {\min}&~{\bf c}^T{\bf z} \\
        {{\text {s.t.}}}\,&~ -\sum_{n=0}^{\frac{N-3}{2}}{\log _2z_n}-\frac{1}{2}\log _2z_{\frac{N-1}{2}}+\frac{N}{2}\log _2\bar{z}\le 0,\label{eq: ine}\\
        &~ {\bf z}>{\bf 0}_{\frac{N+1}{2}},
    \end{align}
    \end{subequations}
    where ${\bf 0}_{\frac{N+1}{2}}$ is a $\frac{N+1}{2}\times 1$ zero vector. Note that \eqref{eq: sim_fixed} is a convex optimization problem and is equivalent to problem $\left(\mathcal{P}_{6.2}\right)$. The global optimal solution of \eqref{eq: sim_fixed} can be obtained by KKT conditions.

    Relaxing ${\bf z}>{\bf 0}_{\frac{N+1}{2}}$ to ${\bf z}\geq{\bf 0}_{\frac{N+1}{2}}$, and defining $\mathbf{v}=\left[ \begin{array}{c}
	\eqref{eq: ine}\\
	-\mathbf{z}\\
    \end{array} \right] $, the KKT conditions for \eqref{eq: sim_fixed} can be expressed as
    \begin{align}
        \mathbf{c}+J_{\mathbf{z}}\left( \mathbf{v} \right) ^T\boldsymbol{\lambda }&=\mathbf{0}_{\frac{N+1}{2}},\label{eq: kkt_1} \\
        \lambda _iv_i&=0,~i=0,\cdots ,\frac{N+1}{2},\label{eq: kkt_2}\\
        \boldsymbol{\lambda }&\ge \mathbf{0}_{\left( \frac{N+1}{2}+1 \right)},    \\
        \mathbf{v}&\le \mathbf{0}_{\left( \frac{N+1}{2}+1 \right)},
    \end{align}
    where $J_{\mathbf{z}}\left( \mathbf{v} \right) =\left[ \mathbf{a}, -\mathbf{I}_{\frac{N+1}{2}} \right] ^T\in \mathbb{R}^{\left(\frac{N+1}{2}+1\right)\times \frac{N+1}{2}}$ with $\mathbf{a}=\frac{1}{\ln 2}\left[ -\frac{1}{z_0},\cdots ,-\frac{1}{z_{\frac{N-3}{2}}},-\frac{1}{2z_{\frac{N-1}{2}}} \right] ^T$ is the Jacobian matrix of $\bf v$, and $\boldsymbol{\lambda }\in \mathbb{R}^{\left(\frac{N+1}{2}+1\right)\times 1}$ is the Lagrangian multipliers vector.

    Note that $z_i\neq 0,~i=0,1,\cdots,\frac{N-1}{2}$, i.e., $v_i\neq 0,~i=1,2,\cdots,\frac{N+1}{2}$. Hence, the Lagrangian multipliers $\lambda_i$ become $\lambda_i = 0,~i=1,2,\cdots,\frac{N+1}{2}$ using \eqref{eq: kkt_2}. Since $c_i\neq 0,~i=0,1,\cdots,\frac{N-1}{2}$, we have $\lambda_0\neq 0$ from \eqref{eq: kkt_1}, and  \eqref{eq: kkt_2} shows $v_0= 0$. In summary, the following three equations are obtained:
    \begin{align}
        c_i&=\frac{\lambda _0}{z_i\ln 2},~i=0,1,\cdots,\frac{N-3}{2},\label{eq: ci_temp}\\
        c_{\frac{N-1}{2}}&=\frac{\lambda _1}{2z_{\frac{N-1}{2}}\ln 2},\label{eq: ce_temp}\\
        \frac{N}{2}\log _2\bar{z}&=\sum_{n=0}^{\frac{N-3}{2}}{\log _2z_n}+\frac{1}{2}\log _2z_{\frac{N-1}{2}}\label{eq: con_eq}.
    \end{align}
    Using \eqref{eq: ci_temp}, \eqref{eq: ce_temp} and \eqref{eq: con_eq}, we have $\lambda _0=\frac{\pi}{6}\ln 2\cdot \bar{z}>0$. Putting $\lambda _0=\frac{\pi}{6}\ln 2\cdot \bar{z}$ into \eqref{eq: ci_temp} and \eqref{eq: ce_temp}, we obtain
    \begin{equation}
        z_i=\bar{z},~i=0,1,\cdots,\frac{N-1}{2}.
    \end{equation}
    The solution satisfies the KKT conditions. Using the definition of $z_i$ and $\bar{z}$, we obtain \eqref{eq: prop_1}. Hence, \textit{Proposition} \ref{prop: fixed} holds.
\end{IEEEproof}

\textit{Proposition} \ref{prop: fixed} reveals that the bit allocation for fixed-point quantization should be distributed equally among all coefficients. Moreover, since $\Tilde{b}_n$ in \eqref{eq: prop_1} is already a non-negative integer solution, the optimal bit allocation for problem $\left(\mathcal{P}_{6.2}\right)$ is also optimal for problem $\left(\mathcal{P}_{6.1}\right)$. Note that the time complexity of solving problem $\left(\mathcal{P}_{6.1}\right)$ is $\mathcal{O}\left(1\right)$ due to the closed-form solution.

\subsubsection{Solution for Floating-Point Quantization}
\label{sec: so_fp}
First, we recall the definition of floating-point numbers. A floating-point number system $\mathbb{F}$ is a subset of real numbers whose elements can be expressed as \cite{higham2022mixed}
\begin{equation}
\label{eq:fp_representation}
    f = \pm k\times \eta^{e-m+1},
\end{equation}
where $\eta = 2$ is the base, the integer $m$ is the mantissa bit, the integer $e$ is the exponent bit within the range $e_{\min}\leq e\leq e_{\max}$, and the integer $k$ is significand satisfying $0\leq k \leq \eta^m - 1$. 

Then, the floating-point quantization model is presented in the following lemma.
\begin{lemma}[Floating-point quantization model \cite{higham2002accuracy,constantinides2021rigorous}]\label{lem: float_quant_error}
    For $b_n$ bit floating-point quantization with $e$ bits of exponent and $m_n$ bits of mantissa, given input filter coefficient $h[n]$, the output $\Hat{h}[n]$ is given by
    \begin{align}
        \Hat{h}[n] &=\boldsymbol{fl}(h[n],b_n)= \boldsymbol{fl}(h[n],[e,m_n])\nonumber \\
        &= h[n]\left(1+\delta_n\right) = h[n] + {h[n]\delta _n}.
    \end{align}
    where $\boldsymbol{fl}(\cdot)$ is the floating-point quantization function, which is the correctly rounded (to nearest) value of inputs, and the relative error $\delta_n$ is a variable with zero mean and ${2^{-2m_n}}/{6}$ variance \cite{constantinides2021rigorous,fang2024statistical}.
\end{lemma}
Compared with \textit{Lemma} \ref{lem: fixed_quant_error}, \textit{Lemma} \ref{lem: float_quant_error} shows that the quantization errors for floating-point arithmetic depend on the inputs, while the fixed-point quantization errors are independent of the inputs. Moreover, since the precision of floating-point quantization, i.e., the variance of the relative error $\delta_n$, depends on the mantissa bit rather than the exponent bit, for simplicity, we assume that different precision floating-point quantizations are regarded as having the same exponent bit, providing sufficient range to prevent overflow and underflow. Therefore, the original quantization bit allocation in problem $\left(\mathcal{P}_{4}\right)$ is transformed into the mantissa bit allocation, allowing us to focus on the mantissa bit in the subsequent paragraphs.

Furthermore, based on \textit{Lemma} \ref{lem: float_quant_error}, \eqref{eq: finite_H} is given by
\begin{align}
    \Hat{H}\left( \omega \right) = &\sum_{n=0}^{\frac{N-3}{2}}2\left(h\left[ n \right]+h[n]\delta _n \right)\cos\left[ \left(\frac{N-1}{2} - n\right)\omega \right]\nonumber\\
    &+ \left(h\left[ \frac{N-1}{2} \right]+h\left[ \frac{N-1}{2} \right]\delta_{\frac{N-1}{2}}\right) \label{eq: finite_H3}.
\end{align}

Similar to the analysis of fixed-point quantization, by assuming the relative errors are independent variables, the objective function \eqref{eq: obj_p2} can be simplified as follows:
\begin{align}
    &\mathbb{E} \left\{ \int_0^{\pi}{\left| \hat{H}\left( \omega \right) -H\left( \omega \right) \right|^2d\omega} \right\}\nonumber\\
    &=\sum_{n=0}^{\frac{N-3}{2}}{\frac{\pi}{3}2^{-2m_n}h^2[n]}+\frac{\pi}{6}2^{-2m_{\frac{N-1}{2}}}h^2\left[ \frac{N-1}{2} \right].\label{eq: obj_p2.3}
\end{align}
where $f_N = \delta_{\frac{N-1}{2}}h\left[ \frac{N-1}{2} \right]$. Then problem $\left(\mathcal{P}_{6}\right)$ can be converted into 
\begin{align*} 
{\left ({{{\mathcal{P}_{6.3}}} }\right)}~{\underset{\{ m_n \}_{n=0}^{N-1}}{\min}  }&~{\eqref{eq: obj_p2.3}}  \\ 
{{\text {s.t.}}}~~\,&~ {2\sum_{n=0}^{\frac{N-3}{2}}{m_n}+m_{\frac{N-1}{2}}\le N\cdot \bar{m},}  \\
&~\ m_n \in \mathbb{Z}_+, \ \forall n = 0, 1, \dots, \frac{N-1}{2}.
\end{align*}



To avoid integer programming, we relax the integer variables ${\bf m}\in \mathbb{Z}_+^{N\times 1}$ in problem $\left(\mathcal{P}_{6.3}\right)$ to the real numbers $\Tilde{\bf m}\in \mathbb{R}^{N\times 1}$ to find a closed-form solution. Specifically, we present the solution of problem $\left(\mathcal{P}_{6.3}\right)$ without integer constraint in the following proposition.
\begin{proposition}[Closed-form solution for the relaxed floating-point quantization problem]\label{prop: float}
    For problem $\left(\mathcal{P}_{6.3}\right)$ without integer constraint, the optimal mantissa bit allocation, i.e., quantization bit allocation, is derived as 
        \begin{align}
        \label{eq: prop_2}
        \Tilde{ m}_n =\bar{m}+\log _2\left( \frac{\left| h\left[ n \right] \right|}{{\rm GM}({\bf h})} \right),~n = 0,1,\cdots,\frac{N-1}{2},
        \end{align}
    where ${\rm GM}({\bf h}) = \left( \prod_{n=0}^{N-1}{\left| h\left[ n \right] \right|} \right) ^{\frac{1}{N}}$.    
\end{proposition}
\begin{IEEEproof}
    The proof is similar to that of \textit{Proposition} \ref{prop: fixed}, which is omitted for conciseness.
\end{IEEEproof}
\textit{Proposition} \ref{prop: float} indicates that the optimal bit $\Tilde{m}_n$ of the $n$-th filter coefficient increases logarithmically with $\left| h\left[ i \right] \right|$ and decreases logarithmically by the geometric mean of the filter coefficients absolute values. Consequently, it can be observed that filter coefficients with larger absolute values require more quantization bits to minimize total quantization loss.

Note that $\Tilde{m}_n$ in \eqref{eq: prop_2} is a real-valued solution, which must be mapped to a non-negative integer. Although a nearest-integer mapping with a greedy criterion could be used, it has high time complexity due to the need to evaluate all possible options. To address this, we propose a low-complexity mapping method that balances bit consumption with quantization loss. Specifically, since the minimum mantissa bit is one \cite{4610935}, i.e., $\Tilde{ m}_n\geq 1$, we have $\bar{m}\geq 1+ \lceil\log _2\left( \frac{{\rm GM}({\bf h})}{\min_{n}\left| h\left[ i \right] \right|} \right)\rceil$. Then, we map the non-integer mantissa bit ($\Tilde{ m}_n\notin \mathbb{Z}$) to $\lceil \Tilde{m}_i \rceil$. If the total bit budget is not met, we need to map the subset of the non-integer mantissa bit to $\lfloor \Tilde{m}_i \rfloor$ rather than $\lceil \Tilde{m}_i \rceil$. Since mapping to $\lfloor \Tilde{m}_i \rfloor$ increases quantization loss, it is crucial to select a good subset. To achieve this, we consider \textit{Lemma} \ref{lem: float_quant_error} to hold for $m_n \in \mathbb{R}$ and propose a trade-off function as follows:
\begin{align}
    K\left( i \right) =\left| \frac{\mathcal{E} _i\left( \tilde{m}_i \right) -\mathcal{E} _i\left( \lfloor \tilde{m}_i \rfloor \right)}{\tilde{m}_i-\lfloor \tilde{m}_i \rfloor} \right|
     =\frac{2^{-2\lfloor \tilde{m}_i \rfloor}-2^{-2\tilde{m}_i}}{\tilde{m}_i-\lfloor \tilde{m}_i \rfloor}c_i, \label{eq: trade-off}
\end{align}
where $\mathcal{E} _i\left( \tilde{m}_i \right) =2^{-2\tilde{m}_i}c_i$ is the mean square quantization error (MSQE) of $i$-th filter coefficient with $\tilde{m}_i$ mantissa bit, $c_i=\frac{\pi}{3}h^2\left[ i \right],~i=0,1,\cdots,\frac{N-3}{2}$, and $ c_{\frac{N-1}{2}}=\frac{\pi}{6}h^2\left[ \frac{N-1}{2} \right]$. Furthermore, \eqref{eq: trade-off} indicates the MSQE increase per unit bit when mapping $\tilde{m}_i$ to $\lfloor \Tilde{m}_i \rfloor$. Thus, we can re-map the values of $\tilde{m}_i$ with the smallest $K(i)$ from $\lceil \Tilde{m}_i \rceil$ to $\lfloor \Tilde{m}_i \rfloor$, achieving a balance between bit consumption and quantization loss. This process is repeated for the next smallest $K(i)$ values until the maximum budget constraint is satisfied.

\begin{algorithm}[t]
\label{alg: LC-FP}
    \SetAlgoNlRelativeSize{0}
    \SetAlgoNlRelativeSize{-1}
    \caption{Low-Complexity Mapping Algorithm}
    \KwIn{$\Bar{m}$, ${\bf h}$, $N$}
    \KwOut{The mantissa bit allocation ${\bf m}$}

    Set $\mathbb{S}={0,1,\cdots,\frac{N-1}{2}}$
    
    \For{$i=0:\frac{N-1}{2}$}{
    Compute $\Tilde{ m}_i$ using \eqref{eq: prop_2} and $m_i=\lceil \Tilde{m}_i \rceil$

    \textbf{if} $\Tilde{m}_i \in \mathbb{Z}$ \textbf{then} $\mathbb{S} = \mathbb{S} - \{i\}$
    }

    Compute the maximum bit $T_{\max}=N\Bar{m}$, and the total bit $T_{\rm total} = 2\sum_{i=0}^{\frac{N-3}{2}}{m_i}+m_{\frac{N-1}{2}}$

    \textbf{for} $i\in \mathbb{S}$ \textbf{do} Compute $K(i)$ using \eqref{eq: trade-off} \textbf{end}


    \While{$T_{\rm total}>T_{\max}$}{
        $i^*=\arg\underset{i\in \mathbb{S}}{\min}~K(i)$

        $m_{i^*} = m_{i^*} - 1$, and $\mathbb{S} = \mathbb{S} - \{i^*\}$
        
        Recompute the total bit $T_{\rm total}$
    }
    
    return ${\bf m}$
\end{algorithm}
The complete procedure of the low-complexity mapping algorithm is detailed in \textit{Algorithm} \ref{alg: LC-FP}. Notably, the while loop in line $8\sim 12$ will always terminate. This is because the total bit consumption always becomes $\sum_{i}\lfloor \Tilde{m}_i \rfloor\leq \sum_{i} \Tilde{m}_i =N\cdot \bar{m}$, i.e., \textit{Algorithm} \ref{alg: LC-FP} always satisfies the maximum bit constraint. Moreover, since the while loop executes at most $\frac{N+1}{2}$ times, the time complexity of \textit{Algorithm} \ref{alg: LC-FP} is $\mathcal{O}\left(N\right)$, lower than that of the PSO-based algorithms proposed and the brute force search method.

\bibliographystyle{IEEEtran}
\bibliography{reference}

\begin{thebibliography}{10}
\providecommand{\url}[1]{#1}
\csname url@samestyle\endcsname
\providecommand{\newblock}{\relax}
\providecommand{\bibinfo}[2]{#2}
\providecommand{\BIBentrySTDinterwordspacing}{\spaceskip=0pt\relax}
\providecommand{\BIBentryALTinterwordstretchfactor}{4}
\providecommand{\BIBentryALTinterwordspacing}{\spaceskip=\fontdimen2\font plus
\BIBentryALTinterwordstretchfactor\fontdimen3\font minus \fontdimen4\font\relax}
\providecommand{\BIBforeignlanguage}[2]{{%
\expandafter\ifx\csname l@#1\endcsname\relax
\typeout{** WARNING: IEEEtran.bst: No hyphenation pattern has been}%
\typeout{** loaded for the language `#1'. Using the pattern for}%
\typeout{** the default language instead.}%
\else
\language=\csname l@#1\endcsname
\fi
#2}}
\providecommand{\BIBdecl}{\relax}
\BIBdecl

\bibitem{10379539}
Z.~Wang \emph{et~al.}, ``A tutorial on extremely large-scale {MIMO} for {6G}: Fundamentals, signal processing, and applications,'' \emph{IEEE Commun. Surveys Tut.}, vol.~26, no.~3, pp. 1560--1605, 2024.

\bibitem{7091863}
K.~Zheng \emph{et~al.}, ``Survey of large-scale {MIMO} systems,'' \emph{IEEE Commun. Surveys Tut.}, vol.~17, no.~3, pp. 1738--1760, 2015.

\bibitem{9165233}
M.~Wang, W.~Fu, X.~He, S.~Hao, and X.~Wu, ``A survey on large-scale machine learning,'' \emph{IEEE Trans. Knowl. Data Eng.}, vol.~34, no.~6, pp. 2574--2594, 2022.

\bibitem{dean2012large}
J.~Dean \emph{et~al.}, ``Large scale distributed deep networks,'' \emph{Proc. Int. Conf. Neural Inf. Process. Syst.}, vol.~25, 2012.

\bibitem{6094235}
D.~M. Kodek, ``{LLL} algorithm and the optimal finite wordlength {FIR} design,'' \emph{IEEE Trans. Signal Process.}, vol.~60, no.~3, pp. 1493--1498, 2012.

\bibitem{7938758}
Y.~Chi and H.~Fu, ``Subspace learning from bits,'' \emph{IEEE Trans. Signal Process.}, vol.~65, no.~17, pp. 4429--4442, 2017.

\bibitem{7478040}
K.~Yu, Y.~D. Zhang, M.~Bao, Y.-H. Hu, and Z.~Wang, ``{DOA} estimation from one-bit compressed array data via joint sparse representation,'' \emph{IEEE Signal Process. Lett.}, vol.~23, no.~9, pp. 1279--1283, 2016.

\bibitem{7307134}
L.~Fan, S.~Jin, C.-K. Wen, and H.~Zhang, ``Uplink achievable rate for massive {MIMO} systems with low-resolution {ADC},'' \emph{IEEE Commun. Lett.}, vol.~19, no.~12, pp. 2186--2189, 2015.

\bibitem{7439790}
J.~Choi, J.~Mo, and R.~W. Heath, ``Near maximum-likelihood detector and channel estimator for uplink multiuser massive {MIMO} systems with one-bit {ADCs},'' \emph{IEEE Trans. Commun.}, vol.~64, no.~5, pp. 2005--2018, 2016.

\bibitem{9043731}
L.~Deng, G.~Li, S.~Han, L.~Shi, and Y.~Xie, ``Model compression and hardware acceleration for neural networks: A comprehensive survey,'' \emph{Proc. IEEE}, vol. 108, no.~4, pp. 485--532, 2020.

\bibitem{sharma2018bit}
H.~Sharma \emph{et~al.}, ``Bit fusion: Bit-level dynamically composable architecture for accelerating deep neural network,'' in \emph{Proc. ACM/IEEE 45th Annu. Int. Symp. Comput. Archit. (ISCA)}.\hskip 1em plus 0.5em minus 0.4em\relax IEEE, 2018, pp. 764--775.

\bibitem{10541876}
X.~Zhang, Y.~Cheng, X.~Shang, and J.~Liu, ``{CRB} analysis for mixed-{ADC} based {DOA} estimation,'' \emph{IEEE Trans. Signal Process.}, vol.~72, pp. 3043--3058, 2024.

\bibitem{10185129}
S.~Yang, Y.~Lai, A.~Jakobsson, and W.~Yi, ``Hybrid quantized signal detection with a bandwidth-constrained distributed radar system,'' \emph{IEEE Trans. Aerosp. Electron. Syst.}, vol.~59, no.~6, pp. 7835--7850, 2023.

\bibitem{7562390}
T.-C. Zhang, C.-K. Wen, S.~Jin, and T.~Jiang, ``Mixed-{ADC} massive {MIMO} detectors: Performance analysis and design optimization,'' \emph{IEEE Trans. Wireless Commun.}, vol.~15, no.~11, pp. 7738--7752, 2016.

\bibitem{7437384}
N.~Liang and W.~Zhang, ``Mixed-{ADC} massive {MIMO},'' \emph{IEEE J. Sel. Areas Commun.}, vol.~34, no.~4, pp. 983--997, 2016.

\bibitem{dettmers2022gpt3}
T.~Dettmers, M.~Lewis, Y.~Belkada, and L.~Zettlemoyer, ``Gpt3. int8 (): 8-bit matrix multiplication for transformers at scale,'' \emph{Proc. Int. Conf. Neural Inf. Process. Syst.}, vol.~35, pp. 30\,318--30\,332, 2022.

\bibitem{8017448}
J.~Choi, B.~L. Evans, and A.~Gatherer, ``Resolution-adaptive hybrid {MIMO} architectures for millimeter wave communications,'' \emph{IEEE Trans. Signal Process.}, vol.~65, no.~23, pp. 6201--6216, 2017.

\bibitem{9966648}
Y.~Xiong, S.~Sun, L.~Liu, Z.~Zhang, and N.~Wei, ``Performance analysis and bit allocation of cell-free massive {MIMO} network with variable-resolution {ADCs},'' \emph{IEEE Trans. Commun.}, vol.~71, no.~1, pp. 67--82, 2023.

\bibitem{9571074}
I.~E. Berman and T.~Routtenberg, ``Resource allocation and dithering of {Bayesian} parameter estimation using mixed-resolution data,'' \emph{IEEE Trans. Signal Process.}, vol.~69, pp. 6148--6164, 2021.

\bibitem{10508306}
M.~Kim, I.-s. Kim, and J.~Choi, ``Meta-heuristic fronthaul bit allocation for cell-free massive {MIMO} systems,'' \emph{IEEE Trans. Wireless Commun.}, vol.~23, no.~9, pp. 11\,737--11\,752, 2024.

\bibitem{Chen_2021_ICCV}
W.~Chen, P.~Wang, and J.~Cheng, ``Towards mixed-precision quantization of neural networks via constrained optimization,'' in \emph{Proc. IEEE/CVF Int. Conf. Comput. Vis. (ICCV)}, October 2021, pp. 5350--5359.

\bibitem{9399174}
W.~Fei, W.~Dai, C.~Li, J.~Zou, and H.~Xiong, ``General bitwidth assignment for efficient deep convolutional neural network quantization,'' \emph{IEEE Trans. Neural Netw. Learn. Syst.}, vol.~33, no.~10, pp. 5253--5267, 2022.

\bibitem{10091800}
M.~Lan, Q.~Ling, S.~Xiao, and W.~Zhang, ``Quantization bits allocation for wireless federated learning,'' \emph{IEEE Trans. Wireless Commun.}, vol.~22, no.~11, pp. 8336--8351, 2023.

\bibitem{nojima2005effects}
Y.~Nojima, K.~Narukawa, S.~Kaige, and H.~Ishibuchi, ``Effects of removing overlapping solutions on the performance of the {NSGA-II} algorithm,'' in \emph{Proc. 3rd Int. Conf. Evol. Multi-Criterion Optim.}\hskip 1em plus 0.5em minus 0.4em\relax Springer, 2005, pp. 341--354.

\bibitem{panda2018performance}
M.~Panda, ``Performance comparison of genetic algorithm, particle swarm optimization and simulated annealing applied to {TSP},'' \emph{Int. J. Appl. Eng. Res.}, vol.~13, no.~9, pp. 6808--6816, 2018.

\bibitem{kodek2005telescoping}
D.~M. Kodek and M.~Krisper, ``Telescoping rounding for suboptimal finite wordlength {FIR} digital filter design,'' \emph{Digit. Signal Process.}, vol.~15, no.~6, pp. 522--535, 2005.

\bibitem{8307077}
N.~Brisebarre, S.-I. Filip, and G.~Hanrot, ``A lattice basis reduction approach for the design of finite wordlength {FIR} filters,'' \emph{IEEE Trans. Signal Process.}, vol.~66, no.~10, pp. 2673--2684, 2018.

\bibitem{schrijver1998theory}
A.~Schrijver, \emph{Theory of linear and integer programming}.\hskip 1em plus 0.5em minus 0.4em\relax John Wiley \& Sons, 1998.

\bibitem{785511}
Y.~Shi and R.~Eberhart, ``Empirical study of particle swarm optimization,'' in \emph{Proc. Congr. Evol. Comput. (CEC)}, vol.~3, 1999, pp. 1945--1950 Vol. 3.

\bibitem{1304846}
A.~Ratnaweera, S.~Halgamuge, and H.~Watson, ``Self-organizing hierarchical particle swarm optimizer with time-varying acceleration coefficients,'' \emph{IEEE Trans. Evol. Comput.}, vol.~8, no.~3, pp. 240--255, 2004.

\bibitem{9680690}
T.~M. Shami \emph{et~al.}, ``Particle swarm optimization: A comprehensive survey,'' \emph{IEEE Access}, vol.~10, pp. 10\,031--10\,061, 2022.

\bibitem{6819057}
R.~Cheng and Y.~Jin, ``A competitive swarm optimizer for large scale optimization,'' \emph{IEEE Trans. Cybern.}, vol.~45, no.~2, pp. 191--204, 2015.

\bibitem{985692}
M.~Clerc and J.~Kennedy, ``The particle swarm - explosion, stability, and convergence in a multidimensional complex space,'' \emph{IEEE Trans. Evol. Comput.}, vol.~6, no.~1, pp. 58--73, 2002.

\bibitem{TRELEA2003317}
I.~C. Trelea, ``The particle swarm optimization algorithm: convergence analysis and parameter selection,'' \emph{Inf. Process. Lett.}, vol.~85, no.~6, pp. 317--325, 2003.

\bibitem{mitra2001digital}
S.~K. Mitra, \emph{Digital Signal Processing: A Computer-Based Approach}.\hskip 1em plus 0.5em minus 0.4em\relax New York, NY, USA: McGraw Hill, 2001.

\bibitem{1083419}
T.~Parks and J.~McClellan, ``Chebyshev approximation for nonrecursive digital filters with linear phase,'' \emph{IEEE Trans. Circuit Theory}, vol.~19, no.~2, pp. 189--194, 1972.

\bibitem{1162497}
D.~Chan and L.~Rabiner, ``Analysis of quantization errors in the direct form for finite impulse response digital filters,'' \emph{IEEE Trans. Audio Electroacoust.}, vol.~21, no.~4, pp. 354--366, 1973.

\bibitem{6457363}
H.~Q. Ngo, E.~G. Larsson, and T.~L. Marzetta, ``Energy and spectral efficiency of very large multiuser {MIMO} systems,'' \emph{IEEE Trans. Commun.}, vol.~61, no.~4, pp. 1436--1449, 2013.

\bibitem{7886292}
J.~Zhang, L.~Dai, Z.~He, S.~Jin, and X.~Li, ``Performance analysis of mixed-{ADC} massive {MIMO} systems over {Rician} fading channels,'' \emph{IEEE J. Sel. Areas Commun.}, vol.~35, no.~6, pp. 1327--1338, 2017.

\bibitem{761034}
R.~Walden, ``Analog-to-digital converter survey and analysis,'' \emph{IEEE J. Sel. Areas Commun.}, vol.~17, no.~4, pp. 539--550, 1999.

\bibitem{6816003}
Q.~Zhang, S.~Jin, K.-K. Wong, H.~Zhu, and M.~Matthaiou, ``Power scaling of uplink massive {MIMO} systems with arbitrary-rank channel means,'' \emph{IEEE J. Sel. Topics Signal Process.}, vol.~8, no.~5, pp. 966--981, 2014.

\bibitem{bekkerman2011scaling}
R.~Bekkerman, M.~Bilenko, and J.~Langford, \emph{Scaling up machine learning: Parallel and distributed approaches}.\hskip 1em plus 0.5em minus 0.4em\relax Cambridge University Press, 2011.

\bibitem{9764884}
C.-Y. Lin, V.~Kostina, and B.~Hassibi, ``Differentially quantized gradient methods,'' \emph{IEEE Trans. Inf. Theory}, vol.~68, no.~9, pp. 6078--6097, 2022.

\bibitem{kolodziej2019suitesparse}
S.~P. Kolodziej \emph{et~al.}, ``The suitesparse matrix collection website interface,'' \emph{J. Open Source Softw.}, vol.~4, no.~35, p. 1244, 2019.

\bibitem{koh2007interior}
K.~Koh, S.-J. Kim, and S.~Boyd, ``An interior-point method for large-scale l1-regularized logistic regression,'' \emph{J. Mach. Learn. Res.}, vol.~8, no. Jul, pp. 1519--1555, 2007.

\bibitem{chang2011libsvm}
C.-C. Chang and C.-J. Lin, ``{LIBSVM}: a library for support vector machines,'' \emph{ACM Trans. Intell. Syst. Technol.}, vol.~2, no.~3, pp. 1--27, 2011.

\bibitem{lecun2002efficient}
Y.~LeCun, L.~Bottou, G.~B. Orr, and K.-R. M{\"u}ller, ``Efficient backprop,'' in \emph{Neural networks: Tricks of the trade}.\hskip 1em plus 0.5em minus 0.4em\relax Springer, 2002, pp. 9--50.

\bibitem{khalil2002nonlinear}
H.~K. Khalil, \emph{Nonlinear Systems}, 3rd~ed.\hskip 1em plus 0.5em minus 0.4em\relax Upper Saddle River, NJ, USA: Prentice-Hall, 2002.

\bibitem{5995137}
J.~Janhunen, T.~Pitkanen, O.~Silven, and M.~Juntti, ``Fixed- and floating-point processor comparison for {MIMO-OFDM} detector,'' \emph{IEEE J. Sel. Topics Signal Process.}, vol.~5, no.~8, pp. 1588--1598, 2011.

\bibitem{boyd2004convex}
S.~Boyd and L.~Vandenberghe, \emph{Convex optimization}.\hskip 1em plus 0.5em minus 0.4em\relax Cambridge university press, 2004.

\bibitem{higham2022mixed}
N.~J. Higham and T.~Mary, ``Mixed precision algorithms in numerical linear algebra,'' \emph{Acta Numer.}, vol.~31, pp. 347--414, 2022.

\bibitem{higham2002accuracy}
N.~J. Higham, \emph{Accuracy and stability of numerical algorithms}, 2nd~ed.\hskip 1em plus 0.5em minus 0.4em\relax Philadelphia, PA, USA: SIAM, 2002.

\bibitem{constantinides2021rigorous}
G.~Constantinides, F.~Dahlqvist, Z.~Rakamari{\'c}, and R.~Salvia, ``Rigorous roundoff error analysis of probabilistic floating-point computations,'' in \emph{Proc. Int. Conf. Comput. Aided Verif.}\hskip 1em plus 0.5em minus 0.4em\relax Springer, 2021, pp. 626--650.

\bibitem{fang2024statistical}
Y.~Fang and L.~Chen, ``Statistical rounding error analysis for random matrix computations,'' \emph{arXiv preprint arXiv:2405.07537}, 2024.

\bibitem{4610935}
``{IEEE} standard for floating-point arithmetic,'' \emph{IEEE Std 754-2008}, pp. 1--70, 2019.

\end{thebibliography}

\end{document}